\documentclass[camera,letterpaper,nomarginnotes,nonarrowgutter]{jpaper}
\usepackage{xspace}
\usepackage{tikz}
\usepackage{xcolor}
\usepackage{soul}
\usepackage{multicol}
\usepackage{multirow}
\usepackage{booktabs}
\usepackage{fancyhdr}
\usepackage{pdfpages}
\usepackage[nomessages]{fp}
\usepackage{tocbibind}
\usepackage{clipboard}

\usepackage[noadjust]{cite}
\usepackage{amsmath,amssymb,amsfonts}
\usepackage{algorithmic}
\usepackage{graphicx}
\usepackage{textcomp}
\usepackage{balance}
\usepackage{mathptmx} 
\usepackage[normalem]{ulem}
\usepackage{url}
\usepackage{marginnote}
\usepackage{enumitem}
\usepackage{ifthen}
\usepackage{caption}
\usepackage{listings}
\usepackage{setspace}
\usepackage{float}
\usepackage[us,12hr]{datetime}
\usepackage[colorinlistoftodos,prependcaption,textsize=small]{todonotes}
\usepackage{xargs}
\usepackage{titletoc}
\usepackage[all]{nowidow}
\usepackage[framemethod=tikz]{mdframed}
\usepackage[most]{tcolorbox}
\tcbuselibrary{breakable}
\tcbuselibrary{hooks}%
\usepackage{flushend}

\usepackage{mathtools}

\usepackage{courier}
\usepackage[italic]{mathastext}

\usepackage{siunitx}
\usepackage{algorithm}
\usepackage{arydshln} %

\usepackage{pifont}
\usepackage{footmisc}
\usepackage{hhline} %
\usepackage[compact]{titlesec} %
\usepackage{footnote}

\usepackage[bookmarks=true,breaklinks=true,letterpaper=true,colorlinks,linkcolor=black,citecolor=blue,urlcolor=black]{hyperref}
\usepackage[resetlabels]{multibib}

\usepackage{circledsteps}

\definecolor{gfored}{rgb}{0.580, 0.050, 0.211}
\definecolor{ao}{rgb}{0.007, 0.520, 0.867}
\definecolor{yt}{rgb}{0.875, 0.568, 1.000}
\definecolor{moegi}{rgb}{0.357, 0.537, 0.188}
\definecolor{jl}{rgb}{1.0, 0.2, 0.8}
\definecolor{brown(web)}{rgb}{0.65, 0.16, 0.16}
\definecolor{bisque}{rgb}{1.0, 0.89, 0.77}

    \usepackage[colorinlistoftodos,prependcaption,textsize=small]{todonotes}

\definecolor{denim}{rgb}{0.08, 0.38, 0.74}
\definecolor{darkolivegreen}{rgb}{0.33, 0.42, 0.18}
\definecolor{dgreen}{rgb}{0.00, 0.75, 0.00}
\definecolor{darkpink}{rgb}{0.88, 0.28, 0.54}
\definecolor{forestgreen}{rgb}{0.0, 0.27, 0.13}
\definecolor{amber}{rgb}{1.0, 0.49, 0.0}
\definecolor{lightyellow}{rgb}{0.980, 0.956, 0.623}
\definecolor{lightblue}{rgb}{0.980, 0.956, 0.623}
\definecolor{joelcolor}{rgb}{1.0, 0.2, 0.6}

\definecolor{dgreen}{rgb}{0.00, 0.5, 0.00}

\newcommand{\heads}[1]{{\noindent\textbf{#1.}\xspace}} %

\newcommand\proposal{TargetCall\xspace}
\newcommand\BaseConv{LightCall\xspace}
\newcommand\lightcallconfigtestes{$5$\xspace}
\newcommand\SimCheck{Similarity Check\xspace}
\newcommand\ltitle{\proposal: Eliminating the Wasted Computation in Basecalling \\ via Pre-Basecalling Filtering\xspace}

\newcommand\target{on-target\xspace}
\newcommand\nontarget{off-target\xspace}

\newcommand\bestfiltersizereduction{$33.31\times$\xspace}

\newcommand\perfoverbasecalling{$3.31\times$\xspace}
\newcommand\accoverbasecalling{$98.88\%$\xspace}
\newcommand\precisionoverbasecalling{$94.71\%$\xspace}
\newcommand\lossoverbasecalling{$1.12\%$\xspace}

\newcommand\recallimpoversigmap{$23.15\%$\xspace}
\newcommand\precisionimpoversigmap{$62.31\%$\xspace}
\newcommand\perfimpoversigmap{$9.72\times$\xspace}
\newcommand\throughputimpoversigmap{$42.08\times$\xspace}
\newcommand\recallimpoveruncalled{$3.09\%$\xspace}
\newcommand\precisionimpoveruncalled{$58.48\%$\xspace}
\newcommand\perfimpoveruncalled{$1.46\times$\xspace}
\newcommand\throughputimpoveruncalled{$1124.03\times$\xspace}

\newcommand\bestperf{$3.31\times$\xspace}
\newcommand\bestrecall{$99.45\%$\xspace}
\newcommand\bestprecision{$96.03\%$\xspace}

\newcommand\worstperf{$2.13\times$\xspace}
\newcommand\worstrecall{$42.57\%$\xspace}
\newcommand\worstprecision{$73.59\%$\xspace}

\newcommand\worstendperf{$2.03\times$\xspace}
\newcommand\bestendperf{$3.00\times$\xspace}

\let\oldmarginnote\marginnote

\renewcommand{\marginnote}[2][rectangle,draw,fill=blue!40,rounded corners]{%
    \oldmarginnote{%
    \tikz \node at (0,0) [#1]{#2};}%
}

\titlespacing*{\section}{0pt}{0ex}{0ex}
\titlespacing*{\subsection}{0pt}{0ex}{0ex}
\titlespacing*{\subsubsection}{0pt}{0ex}{0ex}
\titlespacing\section{1pt}{5pt plus 0.5pt minus 4pt}{5pt plus 0.5pt minus 4pt}
\titlespacing\subsection{1pt}{5pt plus 0.5pt minus 3pt}{5pt plus 0.5pt minus 3pt}
\titlespacing\subsubsection{1pt}{5pt plus 0.5pt minus 2pt}{5pt plus 0.5pt minus 2pt}

\makeatletter
\g@addto@macro{\normalsize}{%
  \setlength{\abovedisplayskip}{2pt plus 1pt minus 1pt}
  \setlength{\belowdisplayskip}{2pt plus 1pt minus 1pt}
  \setlength{\abovedisplayshortskip}{0pt}
  \setlength{\belowdisplayshortskip}{0pt}
  \setlength{\intextsep}{2pt plus 1pt minus 1pt}
  \setlength{\textfloatsep}{3pt plus 1pt minus 1pt}
  \setlength{\dbltextfloatsep}{3pt plus 1pt minus 1pt}
  \setlength{\skip\footins}{4pt plus 1pt minus 1pt}}
\makeatother

\expandafter\def\expandafter\UrlBreaks\expandafter{\UrlBreaks
  \do\a\do\b\do\c\do\d\do\e\do\f\do\g\do\h\do\i\do\j
  \do\k\do\l\do\m\do\n\do\o\do\p\do\q\do\r\do\s\do\t
  \do\u\do\v\do\w\do\x\do\y\do\z\do\A\do\B\do\C\do\D
  \do\E\do\F\do\G\do\H\do\I\do\J\do\K\do\L\do\M\do\N
  \do\O\do\P\do\Q\do\R\do\S\do\T\do\U\do\V\do\W\do\X
  \do\Y\do\Z}

\newcommand{\squishlist}{
 \begin{list}{$\circ$}
  { \setlength{\itemsep}{0pt}
     \setlength{\parsep}{0pt}
     \setlength{\topsep}{0pt}
     \setlength{\partopsep}{0pt}
     \setlength{\leftmargin}{1em}
     \setlength{\labelwidth}{1em}
     \setlength{\labelsep}{0.5em} } }

\newcommand{\squishsublist}{
\begin{list}{$\rightarrow$}
 { \setlength{\itemsep}{0pt}
    \setlength{\parsep}{0pt}
    \setlength{\topsep}{-10em}
    \setlength{\partopsep}{-3pt}
    \setlength{\leftmargin}{1em}
    \setlength{\labelwidth}{1em}
    \setlength{\labelsep}{0.5em} } }

\newcommand{\squishend}{\end{list}}

\let\oldmarginnote\marginnote

\renewcommand{\marginnote}[2][rectangle,draw,fill=blue!40,rounded corners]{%
    \oldmarginnote{%
    \tikz \node at (0,0) [#1]{#2};}%
}

\mdfdefinestyle{graybox}{
    splittopskip=0,%
    splitbottomskip=0,%
    frametitleaboveskip=0,
    frametitlebelowskip=0,
    skipabove=0,%
    skipbelow=0,%
    leftmargin=0,%
    rightmargin=0,%
    innertopmargin=0mm,%
    innerbottommargin=0mm,%
    roundcorner=1mm,%
    backgroundcolor=lightblue,
    hidealllines=true}
    
\mdfdefinestyle{graybox2}{
    splittopskip=0,%
    splitbottomskip=0,%
    frametitleaboveskip=0,
    frametitlebelowskip=0,
    skipabove=0,%
    skipbelow=0,%
    leftmargin=0,%
    rightmargin=0,%
    innertopmargin=2mm,%
    innerbottommargin=2mm,%
    roundcorner=2mm,%
    backgroundcolor=lightblue,
    hidealllines=true}

\makeindex

\newcommand{\affilETH}[0]{\small {$^1$}}
\newcommand{\affilBilkent}[0]{\small {$^2$}}
\title{\ltitle} 
\author{\vspace{-17pt}\\%
\fontsize{11}{12}\selectfont%
{Meryem Banu Cavlak\affilETH{}}\quad%
{Gagandeep Singh\affilETH{}}\quad%
{Mohammed Alser\affilETH{}}\quad%
{Can Firtina\affilETH{}}\quad%
{Jo\"el Lindegger\affilETH{}}
\vspace{-1pt}\\%
\fontsize{11}{12}\selectfont%
{Mohammad Sadrosadati\affilETH{}}\quad%
{Nika Mansouri Ghiasi\affilETH{}}\quad%
{Can Alkan\affilBilkent{}}\quad%
{Onur Mutlu\affilETH{}}%
\vspace{-1pt}\\%
{\fontsize{10}{11}\selectfont
\affilETH\emph{ETH Z\"urich}%
\qquad\quad%
\quad%
\affilBilkent\emph{Bilkent University}%
}
\vspace{-16pt}\vspace{0.3em}}

\newcites{supp}{Supplementary References}
\newcites{rev}{Revision References}

\begin{document}
\bstctlcite{IEEEexample:BSTcontrol}
\maketitle
\thispagestyle{plain}
\pagestyle{plain}

\begin{abstract}
Basecalling is an essential step in nanopore sequencing analysis where the raw signals of nanopore sequencers are converted into nucleotide sequences, i.e., reads. State-of-the-art basecallers employ complex {deep learning models} to {achieve} high basecalling accuracy. This makes basecalling computationally inefficient and memory-hungry, bottlenecking the entire genome {analysis pipeline}. However, for many applications, the majority of reads do no match the reference genome of interest (i.e., target reference) and thus are discarded in later steps in the genomics pipeline, wasting the basecalling computation. To overcome this issue, we propose \proposal{}, the first pre-basecalling filter to eliminate the wasted computation in basecalling. \proposal{}'s key idea is to discard reads that will not match the target reference (i.e., \nontarget{} reads) prior to basecalling. \proposal{} consists of two main components: (1)~\BaseConv{}, a lightweight neural network basecaller that produces noisy reads; and (2)~\SimCheck{}, which labels each of these noisy reads as \target{} or \nontarget{} by matching them to {the} target reference. Our thorough experimental evaluations show that \proposal{} 1) improves the end-to-end basecalling runtime performance {of} the state-of-the-art basecaller by \perfoverbasecalling{} while maintaining high ($98.88\%$) {recall} in keeping \target{} reads, 2) maintains high accuracy in downstream analysis, and 3) achieves better runtime performance, throughput, recall, precision, and generality compared to prior works. \proposal{} is available at \href{https://github.com/CMU-SAFARI/TargetCall}{https://github.com/CMU-SAFARI/TargetCall}.
\end{abstract}

\section{Introduction} \label{sec:introduction}

Genome sequencing, which determines the nucleotide sequence of an organism's genome, plays a pivotal role in enabling many medical and scientific advancements~\cite{alkan2009personalized, ashley2016towards, chin2011cancer, ellegren2014genome, alvarez2017next}. Modern sequencing technologies produce increasingly large amounts of genomic data at low cost~\cite{wang2021nanopore}. Leveraging this genomic data requires fast, efficient, and accurate analysis tools.

Current sequencing machines are unable to determine an organism’s genome as a single contiguous sequence \cite{alser2021technology}. Instead, they sequence fragments of a genome, called \emph{reads}. The length of the reads depends on the sequencing technology and significantly affects the performance (i.e., speed or runtime) and accuracy of genome analysis. The use of \emph{long reads} can provide higher accuracy and performance on many genome analysis steps~\cite{treangen2011repetitive, firtina2016genomic, alkan2010limitations, lu2016oxford, magi2018nanopore, firtina_blend_2023}.

Nanopore sequencing technology is one of the most prominent and widely-used long read sequencing technologies~\cite{wang2021nanopore, branton2008potential, gong2019ultralong, jain2018nanopore, logsdon2020long, amarasinghe2020opportunities}. Nanopore sequencing relies on measuring the change in the electrical current when a nucleic acid molecule (DNA or RNA) passes through a pore of nanometer size~\cite{zhang2021realtime}. The measurement of the electrical current, called a \emph{raw signal}, is converted to a nucleotide sequence, called a \emph{read}, with a step called \emph{basecalling}~\cite{wang2021nanopore, alser2021technology, wick2019performance, pages-gallego_comprehensive_2023, alser2022molecules, wan2021beyond}. Basecalling commonly uses computationally expensive deep neural network (DNN)-based architectures to achieve high basecalling accuracy~\cite{senol2019nanopore, rang2018squiggle, singh_rubicon_2024}, which makes basecalling a computational bottleneck for genome analysis that consumes up to $84.2\%$ of total execution time in genome analysis pipeline~\cite{bowden2019sequencing}. However, the majority of this computation is wasted for genome sequencing applications that do not require the majority of the basecalled reads. For example, in SARS-CoV-2 genome assembly, 96\% of the total runtime is spent on basecalling, even though $\geq$99\% of the basecalled reads are not required after basecalling because they are not coming from the reference genome that is targeted by the application~\cite{dunn2021squigglefilter}. Therefore, it is important to eliminate wasted computation in basecalling. 

Our \textbf{goal} in this work is to eliminate the wasted computation when basecalling the entire read while maintaining high accuracy and applicability to a wide range of genome sequencing applications. To this end, we propose \proposal{}, the \emph{first} pre-basecalling filter. \proposal{} is based on the key observation that typically the reason for discarding basecalled reads is that they do not match some \emph{target reference} (e.g., a reference genome of interest)~\cite{grumaz2016nextgeneration, dunn2021squigglefilter}. We call these \emph{\nontarget{}} reads. Our \textbf{key idea} is to filter out \nontarget{} reads before basecalling by analyzing the \emph{entire} read with a highly accurate and high-performance \emph{pre-basecalling filter} to eliminate the wasted computation in basecalling of \nontarget{} reads.

Prior works in targeted sequencing ~\cite{kovaka2021targeted, zhang2021realtime, dunn2021squigglefilter, bao2021squigglenet,firtina_rawhash_2023,lindegger_rawalign_2023,firtina_rawhash2_2024} propose {\emph{adaptive sampling}} techniques to discard \nontarget{} reads during sequencing to better utilize sequencers. Sequencers provided by Oxford Nanopore Technologies (ONT) can enable adaptive sampling with a feature known as \emph{Read Until}~\cite{kovaka2021targeted, loose2016realtime}. ONT sequencers that support Read Until can selectively remove a read from the nanopore while the read is being sequenced. This requires a method to identify which reads are \nontarget{} for further downstream analysis to decide which reads to remove from the nanopore. The state-of-the-art adaptive sampling methods can be classified into three groups on their methodology to label the read. The first group converts the target reference into a reference raw signal and performs raw signal-level alignment~\cite{zhang2021realtime, dunn2021squigglefilter, loose2016realtime}. The second group generates noisy sequence representations of the raw signal to compare them with the target reference~\cite{kovaka2021targeted, payne2020readfish}. The third group of works utilizes neural network classifiers to label the sequences~\cite{bao2021squigglenet, noordijk_baseless_2023}. We provide a detailed background on different adaptive sampling approaches in Supplementary Section~\ref{sec:adaptivesampling}.

Even though the labeling techniques of adaptive sampling can be repurposed for pre-basecalling filtering, the adaptive sampling problem is different from pre-basecalling filtering for three main reasons. First, in adaptive sampling, reads must be labeled during sequencing, requiring only the initial portion of the raw signal to classify reads as \nontarget{} or \target{}. Analyzing a sub-region or raw signals in adaptive sampling methods often leads to low recall ($77.5\%-90.40\%$)~\cite{kovaka2021targeted, zhang2021realtime} or poor basecalling~\cite{zhang2021realtime,payne2020readfish}, meaning they can falsely reject many \target{} reads. In contrast, a pre-basecalling filter can utilize the entire raw signal after the read is fully sequenced, enabling more accurate classification. Second, adaptive sampling has practical limitations, such as the risk of nanopores becoming blocked after a few seconds of sequencing~\cite{munro_icarust_2024}, which limits the effectiveness of read ejection. A pre-basecalling filter addresses these limitations by processing the whole signal of all reads, even when adaptive sampling is not feasible. Third, some adaptive sampling methods require re-training classifiers for each different application and target reference~\cite{bao2021squigglenet}, while a pre-basecalling filter can be applied without needing re-training, making it more flexible across different use cases. We conclude that pre-basecalling filtering is orthogonal to adaptive sampling and can complement it. Even when adaptive sampling is used to reject reads early, any remaining reads still need to be basecalled, and a pre-basecalling filter can further improve accuracy and efficiency by processing the entire signal of these remaining reads. This makes the pre-basecalling filter a versatile solution that can be applied both independently and in conjunction with adaptive sampling approaches. \proposal{} aims to overcome the challenges of state-of-the-art methods by utilizing the entire raw signal for classification, making it a widely applicable solution.

\proposal{} consists of two main components: 1)~\emph{\BaseConv{}}, a light-weight basecaller with a simple neural network model that outputs erroneous (i.e., noisy) reads with high performance; and 2)~\emph{\SimCheck{}} to compute the similarity of the noisy read to the target reference where the similarity is determined by conventional read mapping pipeline. \BaseConv{}'s model is \bestfiltersizereduction{} smaller than the state-of-the-art basecaller, Bonito's model~\cite{bonito}. This reduction in the model improves the basecalling speed substantially with a small ($4.85\%$) reduction in basecalling accuracy. Although reducing the basecalling accuracy might cause \BaseConv{} to be not applicable to some of the downstream analyses that require high basecalling accuracy~\cite{frei2021ultralong}, it is sufficient for \SimCheck{} to perform pre-basecalling filtering.  We use the state-of-the-art read mapper minimap2~\cite{li_minimap2_2018} for \SimCheck{}. \proposal{} overcomes all three limitations of prior methods. First, \SimCheck{}'s high accuracy enables \proposal{}'s accuracy to be significantly higher than prior adaptive sampling approaches. Second, \BaseConv{}'s performance is independent of the target reference size, which enables \proposal{} to be applicable to target reference sizes for which prior works were inapplicable. Third, unlike prior approaches that require re-training the network for each application and target reference, \BaseConv{} does not need to be re-trained. 

\textbf{Key Results.} We evaluate the performance and accuracy impact of \proposal{} on the state-of-the-art basecaller Bonito~\cite{bonito}, and compare \proposal{} with two state-of-the-art adaptive sampling methods, UNCALLED~\cite{kovaka2021targeted} and Sigmap~\cite{zhang2021realtime}, repurposed as pre-basecalling filters. \proposal{} 1)~improves the end-to-end runtime performance (i.e., runtime of all the steps, including basecalling and read mapping, used in an analysis) by \perfoverbasecalling{} over Bonito, 2) precisely filters out \precisionoverbasecalling{} of the \nontarget{} reads, and 3) maintains high recall in keeping \target{} reads with \accoverbasecalling{} recall. We show that \proposal{} provides high accuracy in a specific downstream analysis that aims to estimate the relative abundance (RA) of organisms in a given sample even after losing \lossoverbasecalling{} of \target{} reads. We demonstrate that \proposal{} improves 1) runtime performance by \perfimpoveruncalled{}/\perfimpoversigmap{}, 2) throughput by \throughputimpoveruncalled{}/\throughputimpoversigmap{}, 3) recall by +\recallimpoveruncalled{}/+\recallimpoversigmap{}, and 4) precision by +\precisionimpoveruncalled{}/+\precisionimpoversigmap{}
over prior works UNCALLED/Sigmap while requiring much less peak memory (on average $5.76\times$) and maintaining scalability to longer target references.

This paper makes the following contributions:
\begin{itemize}
\item We introduce the problem of pre-basecalling filtering that aims to classify reads utilizing the entire raw signal information.
\item We introduce the first pre-basecalling filter that eliminates the wasted computation in basecalling by leveraging the fact that the majority of reads are discarded after basecalling. 
\item We propose \BaseConv{}, a lightweight neural network model that significantly increases the performance of basecalling with minor reductions in basecalling accuracy.
\item \proposal{} provides larger runtime performance and accuracy benefits compared to the state-of-the-art adaptive sampling works for basecalling.
\item To aid research and reproducibility, we freely open source our implementation of \proposal{} at \href{https://github.com/CMU-SAFARI/TargetCall}{https://github.com/CMU-SAFARI/TargetCall}.
\end{itemize}

\section{Materials and Methods} \label{sec:methods}

Our goal in this work is to eliminate the wasted computation in basecalling using an accurate pre-basecalling filtering technique. To this end, we propose \proposal{}, that can perform pre-basecalling filtering in \textit{all} genome sequencing applications accurately and efficiently without any additional overhead. To our knowledge, \proposal{} is the first pre-basecalling filter that is applicable to a wide range of use cases and makes use of entire raw signal information to classify nanopore raw signals. \proposal{}'s key idea is to quickly filter out \nontarget{} reads (i.e., reads that are dissimilar to the \emph{target reference}.) before the basecalling step to eliminate the wasted computation in basecalling. We present the high-level overview of \proposal{} in Section~\ref{subsec:overview}, and explain its components in Sections~\ref{subsec:baseconv} and~\ref{subsec:simcheck}.

\subsection{High Level Overview}\label{subsec:overview}

\begin{figure}[h]
  \centering
  \includegraphics[width=0.95\linewidth]{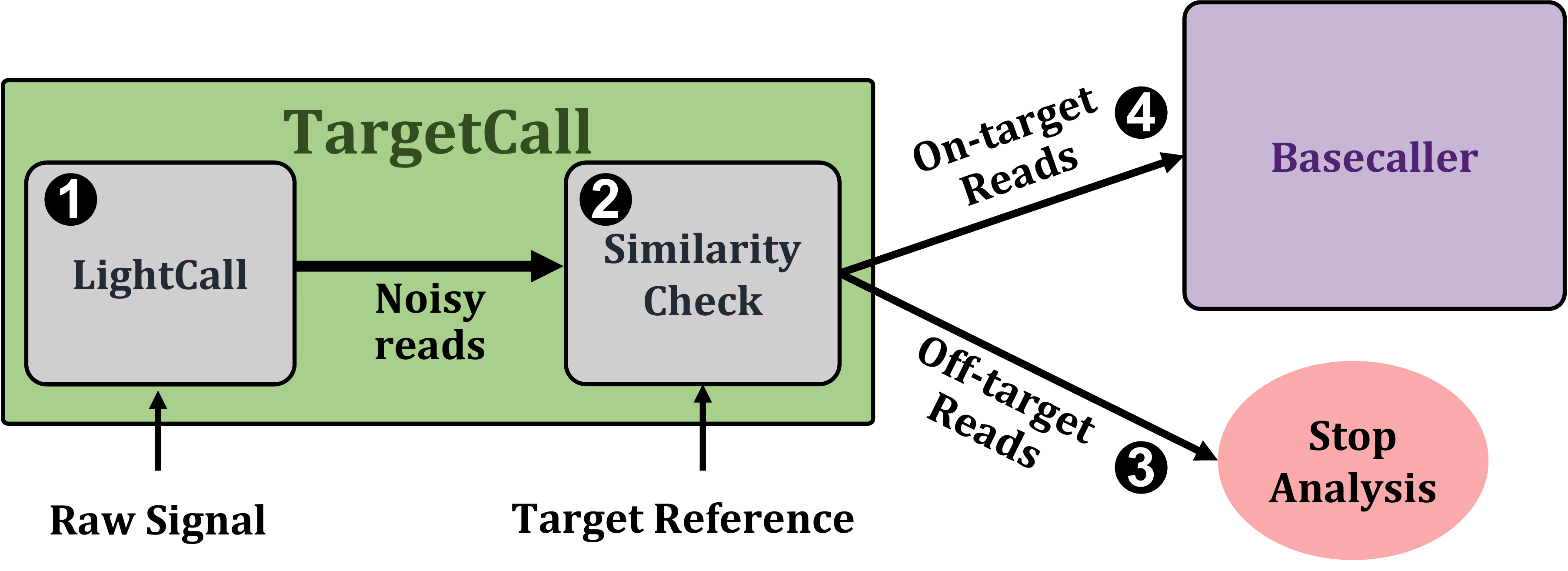}
    \caption{{High-level overview of \proposal{}.}}
  \label{fig:keyidea}
\end{figure}

Figure~\ref{fig:keyidea} {shows \proposal{}'s} workflow. First, \proposal{} performs noisy basecalling on the raw signal using \BaseConv{}~(1). The output sequence of \BaseConv{} is highly accurate but erroneous compared to the reads basecalled using state-of-the-art basecallers. Second, \SimCheck{} compares the noisy read of \BaseConv{} to the target reference for labeling the read as an \target{} or \nontarget{} read~(2). \proposal{} stops the analysis of \nontarget{} reads by removing them from the pipeline~(3), whereas the analysis of the \target{} reads continues with basecalling following the usual genomics pipeline to maintain basecalling accuracy\footnote{Basecalling accuracy is calculated as the ratio of the number of correctly identified bases to the total number of bases in the read, expressed as a percentage.}~(4). 

The choice of the target reference depends on the specific genome sequencing application. The size of the target reference is a major constraint in most prior works, limiting the generality of the prior approaches. We design \proposal{} such that its runtime performance scales well with the size of the target reference so that it is applicable to any genome sequencing application. We achieve our design goal because 1) the performance of \BaseConv{} is independent of the size of the target reference, and 2) the performance of \SimCheck{} module scales well with the increasing target reference size. 

\subsection{\BaseConv{}}\label{subsec:baseconv}

\BaseConv{}, the first component of \proposal{}, is a lightweight neural network-based basecaller that produces noisy reads. Although \BaseConv{} is not designed to be as accurate as the state-of-the-art basecallers, its combination with \SimCheck{} is effective in determining if the read is an \target{} read with respect to the target reference. We develop \BaseConv{} by modifying the state-of-the-art basecaller Bonito's architecture in three ways: 1) reducing the channel sizes of convolution layers, 2) removing the skip connections, and 3) reducing the number of basic convolution blocks. Prior work~\cite{singh_rubicon_2024} shows that Bonito's model is over-provisioned, and we can maintain very high accuracy with reduced model sizes. 
Following prior work's insight, we generate different neural network models by pruning the channel sizes of convolution and convolution blocks. The specific \BaseConv{} configurations are tested and designed based on our intuition. We select the neural network architecture for \BaseConv{} that balances basecalling accuracy with the pre-basecalling filtering performance.

\begin{figure}[h]
  \centering
  \includegraphics[width=0.5\linewidth]{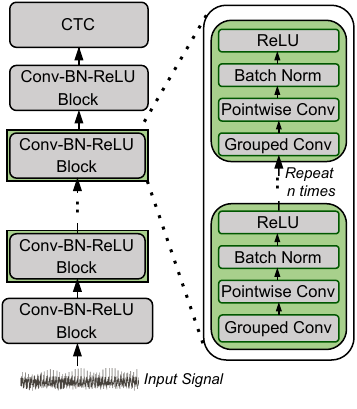}
  \caption{Overview of \BaseConv{} architecture.}
  \label{fig:hc_filter}
  \end{figure}

Figure~\ref{fig:hc_filter} shows the architecture of \BaseConv{}. Each block consists of grouped 1-dimensional convolution and pointwise 1-dimensional convolution. The convolution operation is followed by batch normalization (Batch Norm)~\cite{ioffe2015batch} and a rectified linear unit (ReLU)~\cite{agarap2018deep} activation function. The final output is passed through a connectionist temporal classification (CTC)~\cite{graves2006connectionist} layer to produce the decoded sequence of nucleotides (A, C, G, T). The CTC layer acts as the loss function by providing the correct alignment between the input and the output sequence. The CTC loss allows the model to handle the variable-length sequences by aligning the model's output to the target sequence while ignoring the blank or padding symbols. Our \BaseConv{} architecture is composed of 18 convolution blocks containing $\sim$292 thousand model parameters ($\sim$33.35$\times$ lower parameters than Bonito). 

Modern deep learning-based basecallers~\cite{singh_rubicon_2024, wick2019performance, neumann2022rodan, konishi2021halcyon, xu2021fast,lou2020helix, perevsini2021nanopore, pages-gallego_comprehensive_2023, ambernas_zhang2021automated} incorporate skip connections to help mitigate the vanishing gradient and saturation problems~\cite{anupama2022SynthNet}. Removing skip connections has a higher impact on basecalling accuracy. However, adding skip connections introduces the following three issues for performance~\cite{shahroodi2023swordfish,singh_rubicon_2024}. First, the skip connections increase the data lifetime. The layers whose activations are reused in subsequent layers must wait for this activation reuse (or buffer the activations in memory) before accepting new input and continuing to compute.  This leads to high resource and storage requirements due to data duplication. Second, skip connections introduce irregularity in neural network architecture as these connections span across non-adjacent layers.  Third, skip connections require additional computation to adjust the channel size to match the channel size at the non-consecutive layer's input. Therefore, we remove the skip connections as we can tolerate lower accuracy of \BaseConv{} to improve the performance of \proposal{}.

\BaseConv{} works by splitting a long read in raw signal format (e.g., millions of samples) into multiple smaller chunks (e.g., thousands of samples per chunk) and basecalling these chunks. The CTC layer assigns a probability for all possible labels (i.e., A, C, G, T) at each sequence position for a chunk. The label with the highest probability is selected as the final output for a sequence position. \BaseConv{} merges the outputs of each position to produce the basecalled chunk and merges the basecalled chunks to output the basecalled read. Since \BaseConv{}'s algorithm is independent of the target reference, \BaseConv{}'s performance does not depend on the target reference length.

\subsection{\SimCheck{}}\label{subsec:simcheck}

After \BaseConv{} outputs the noisy read that approximately represents the raw signal, \SimCheck{} compares the noisy read to the target reference. For this task, we use a procedure common in genome analysis, known as sequence alignment. Sequence alignment computes the similarity between a read and a reference. To provide a scalable and fast solution for \SimCheck{}, we use minimap2, a well-optimized sequence aligner, as it is~\cite{li_minimap2_2018}. \SimCheck{} labels a read that is similar to the target reference, i.e., a read that maps to the target reference, as \target{}. With the efficient index structure of minimap2, the performance of sequence alignment is almost independent of the length of the target reference genome. Hence, \proposal{} is scalable to large target reference genomes of Gbp (i.e., Giga base pair) length. 

The accuracy of the computed sequence alignments is not high enough to represent the true alignment between the read and the target reference since the reads computed by \BaseConv{} are noisy. However, the labeling accuracy of \SimCheck{} for determining the \target{}/\nontarget{} reads is high enough to provide recall up to \bestrecall{} in filtering reads. The minor inaccuracy of \SimCheck{} can be compensated with the high sequencing depth-of-coverage, the average number of reads that align to a genomic region, required for confident genome sequence analysis~\cite{quail2012tale, levy2016advancements, hu2021nextgeneration, sims2014sequencing}.

\section{Results}
\subsection{Experimental Setup}\label{subsec:evalmethodology}
\textbf{Evaluated Use Cases.} To show the applicability of \proposal{}, we evaluate it on use cases with varying target reference sizes without compromising accuracy. We describe three use cases we use to evaluate \proposal{}: (1) Covid Detection, (2) Sepsis Detection, and (3) Viral Detection. All three use cases contain a significant fraction of \nontarget{} reads that are eliminated using \proposal{} to show the benefits of pre-basecalling filtering.

\textbf{Covid Detection.} The first use case aims to accept reads coming from a small target reference. We choose SARS-CoV-2 detection as a sample biological application where the goal is to detect the reads coming from a SARS-CoV-2 reference genome ($\sim$30 Kbp) from a sample taken from a human and filter out the human reads in the sample as performed by~\cite{dunn2021squigglefilter,geoghegan2021use}. 

\textbf{Sepsis Detection.} The second use case aims to filter reads when the target reference is large ($\sim$3 Gbp) to show \proposal{} is applicable to use cases with large target references. We choose sepsis detection, as used in~\cite{grumaz2016nextgeneration, celik2022diagnosis, sands2021characterization}, as a sample biological application where the goal is to delete the human reads from a human sample. Since the bacteria that is causing the disease is unknown, we cannot search for reads coming from a specific bacterial target. Instead, we apply \proposal{} to filter out reads similar to the target reference. 

\textbf{Viral Detection.} The third use case aims to filter reads when the target reference contains a collection of reference genomes to show \proposal{} can correctly filter reads when the sample and the target reference have a wide variety of species. This is to test the specificity of \proposal{} in filtering reads when the \target{} and \nontarget{} reads resemble each other more compared to previous use cases. We choose disease-causing viral read detection as a sample biological application where the goal is to detect the viral reads from a metagenomic sample of bacterial \& viral reads, and the target reference contains a collection of viral reference genomes as demonstrated by~\cite{mokili2012metagenomics}. 

\textbf{Evaluation System.} We use NVIDIA TITAN V to train and evaluate \BaseConv{} and Bonito baseline. For our evaluations, we increase the batch size maximally such that the entire GPU memory is occupied (Sections \ref{sec:readaccuracy}, \ref{sec:endaccuracy}, \ref{sec:performance}, \ref{sec:endperformance}). We use NVIDIA A100 to evaluate \proposal{} against prior work. We evaluate UNCALLED and Sigmap on a high-end server (AMD EPYC 7742 CPU with 1TB DDR4 DRAM). For our evaluations, we optimize the number of threads (128) for UNCALLED/Sigmap and the batch size (128) for \proposal{} such that all tools have the minimum execution time (Section \ref{subsec:comparepriorwork}). We use the state-of-the-art read mapper, minimap2~\cite{li_minimap2_2018} for the \SimCheck{} module of \proposal{} with -a and -x map-ont flags. The -a flag is used to compute sequence alignments, and \texttt{-x map-ont} is used to configure minimap2 parameters for ONT data.

\textbf{Training Setting.} We use the publicly available ONT dataset~\cite{bonito} sequenced using MinION Flow Cell (R9.4.1)~\cite{MinIONFlowCell} for the training and validation. The neural network weights are updated using Adam~\cite{kingma2014adam} optimizer with a learning rate of  2$e^{-3}$, a beta value of 0.999, a weight decay of 0.01, and an epsilon of 1$e^{-8}$.

\textbf{Baseline Techniques.} We evaluate \proposal{}'s runtime performance and accuracy as a pre-basecalling filter by integrating it as a pre-basecalling filter to Bonito~\cite{bonito}, which is one of the official basecalling tools developed by ONT. In Section~\ref{subsec:comparepriorwork}, we evaluate two state-of-the-art, non-machine learning-based adaptive sampling methods, UNCALLED~\cite{kovaka2021targeted} and Sigmap~\cite{zhang2021realtime} repurposed as pre-basecalling filters to compare against \proposal{}. We repurposed UNCALLED and Sigmap by using their classification methods to classify the reads as \target{}/\nontarget{} before the basecalling step without any change in their implementations. We identify the ground truth \target{} and \nontarget{} by basecalling the reads using a high accuracy Bonito model followed by performing minimap2.
We do not evaluate \proposal{} against other adaptive sampling methods such as SquiggleNet~\cite{bao2021squigglenet} that cannot be trivially used as pre-basecalling filters. 

\textbf{\BaseConv{} Configurations Evaluated.} To determine the final architecture of \proposal{}, we test \lightcallconfigtestes{} different \BaseConv{} configurations. Table~\ref{tab:baseconv-config} lists the \BaseConv{} configurations evaluated. 

\begin{table}[h]
\centering
\caption{Different \BaseConv{} configurations.}\label{tab:baseconv-config}
\begin{tabular}{@{}lrr@{}}\toprule
\textbf{Model} & \textbf{Number of} & \textbf{Model}\\
\textbf{Name} & \textbf{Parameters} & \textbf{Size (MB)} \\\midrule
Bonito	&	9,739K & 37.14 \\ \midrule
$LC_{Main\times2}$	 & 565K & 2.16 \\\midrule
$LC_{Main}$ &	292K & 1.11 \\\midrule
$LC_{Main/2}$ &	146K & 0.55 \\\midrule
$LC_{Main/4}$ &	52K & 0.19 \\\midrule
$LC_{Main/8}$ &	21K & 0.07 \\\midrule
\end{tabular}
\end{table}

\textbf{Evaluated Datasets.} We sampled $287,767$ reads from prior work~\cite{zook2019open, cadde, wick2019performance} and simulated $35,000$ reads using DeepSimulator~\cite{li2018deepsimulator, li2020deepsimulator} to evaluate \proposal{}. We use four reference genomes to evaluate \proposal{} on three different genome sequencing applications. The details of the exact read datasets and reference genomes used to produce all our results can be found in Supplementary Section~\ref{sec:dataset}.

\textbf{Evaluation Metrics.} We evaluate \proposal{} using five different metrics: 1) filtering accuracy, 2) basecalling accuracy, 3) relative abundance (RA) estimation accuracy, 4) basecalling execution time, and 5) end-to-end execution time. For the basecalling execution time, we compare the wall-clock time spent on pre-basecalling filtering followed by basecalling of the reads that are accepted by the filter and conventional basecalling.
For the end-to-end execution time, we compare the wall-clock time spent on the entire genome analysis pipeline of basecalling, read mapping, and variant calling with and without the use of pre-basecalling filtering. The index generation time of minimap2 is excluded from end-to-end execution time, as this is a one-time task per reference genome. When comparing \proposal{} with UNCALLED and Sigmap in terms of the end-to-end execution time, we acknowledge that \proposal{} benefits from hardware acceleration as it uses GPUs while the other tools use CPUs. Implementing UNCALLED and Sigmap on GPUs could further improve their speed performance.

We evaluate the filtering accuracy of \proposal{} by computing its precision and recall. We define precision as the number of reads that \proposal{} correctly labels as \target{}, divided by the total number of reads that \proposal{} labels as \target{}. We define recall as the number of reads that \proposal{} correctly labels as \target{}, divided by the overall number of \target{} reads in the dataset. The ground truth \target{} reads are determined by the conventional pipeline of basecalling with Bonito and read mapping. An ideal pre-basecalling filter should have 100\% recall to maintain accuracy in the downstream analyses and 100\% precision to provide the maximum runtime performance improvement possible.

For basecalling accuracy, we use Bonito's training and evaluation procedure to extract the median identity as basecalling accuracy~\cite{bonito}. For relative abundance (RA) accuracy, we calculate the difference in relative abundances of viral species after 1) pre-basecalling filtering and 2) conventional basecalling. We compute how much RA deviates from the true RAs after pre-basecalling filtering. RA is defined as the proportion of reads corresponding to a particular species relative to the total number of reads. The true RA refers to the RA obtained after conventional basecalling, which is considered the benchmark. Equation~(\ref{eq:1}) provides the calculation of the deviation in RAs where $TC\_RA_{i}$ is the RA of species $i$ after \proposal{}; and $B\_RA_{i}$ is the RA of species $i$ after conventional basecalling with Bonito.

\begin{equation} \label{eq:1}
\text{RA Deviation} = \sum_{for\ each\ species\ i} 100 * \frac{\left| TC\_RA_{i} - B\_RA_{i} \right|}{B\_RA_{i}}
\end{equation}

\subsection{Filtering Accuracy} \label{sec:readaccuracy}
\begin{figure}[h]
  \centering
  \includegraphics[width=0.95\linewidth]{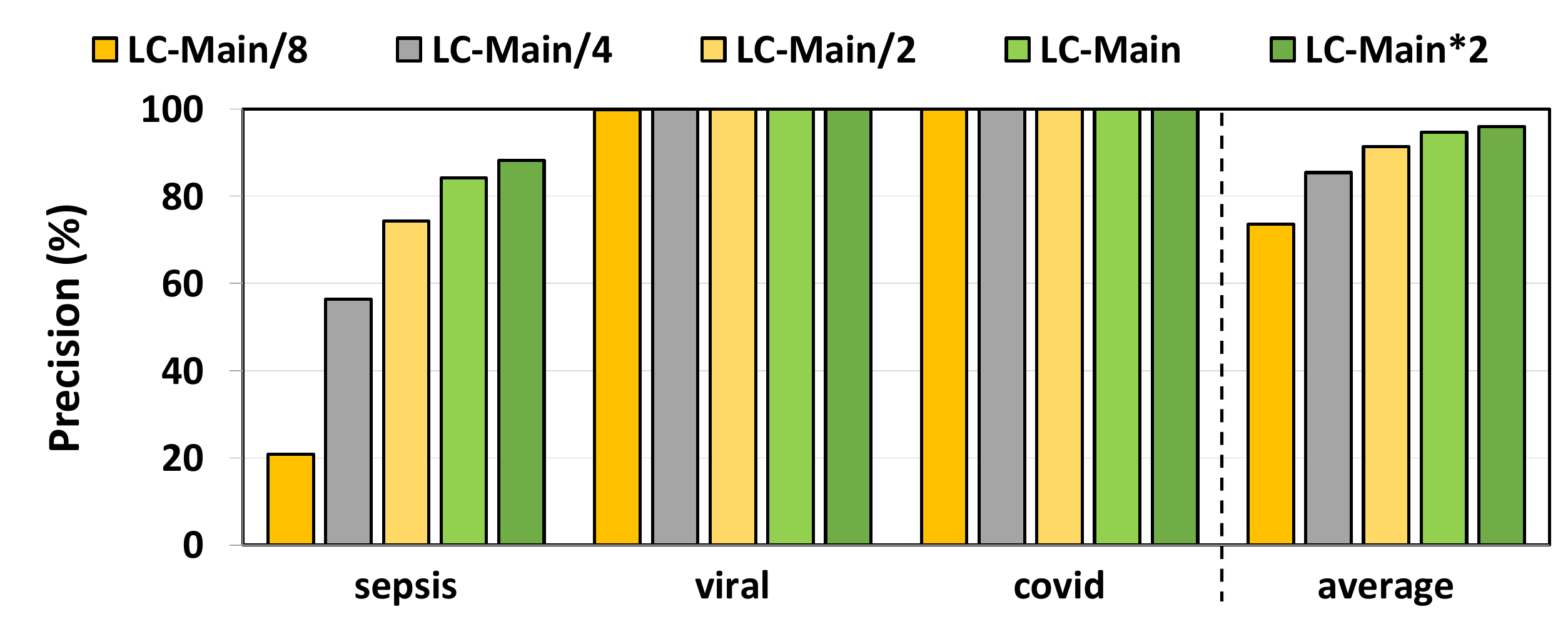}
  \caption{Precision {for our evaluated use-cases while using \proposal{} with different} \BaseConv{} configurations.}\vspace{-3pt}
  \label{fig:precision}
\end{figure}

\begin{figure}[h]
  \centering
  \includegraphics[width=0.95\linewidth]{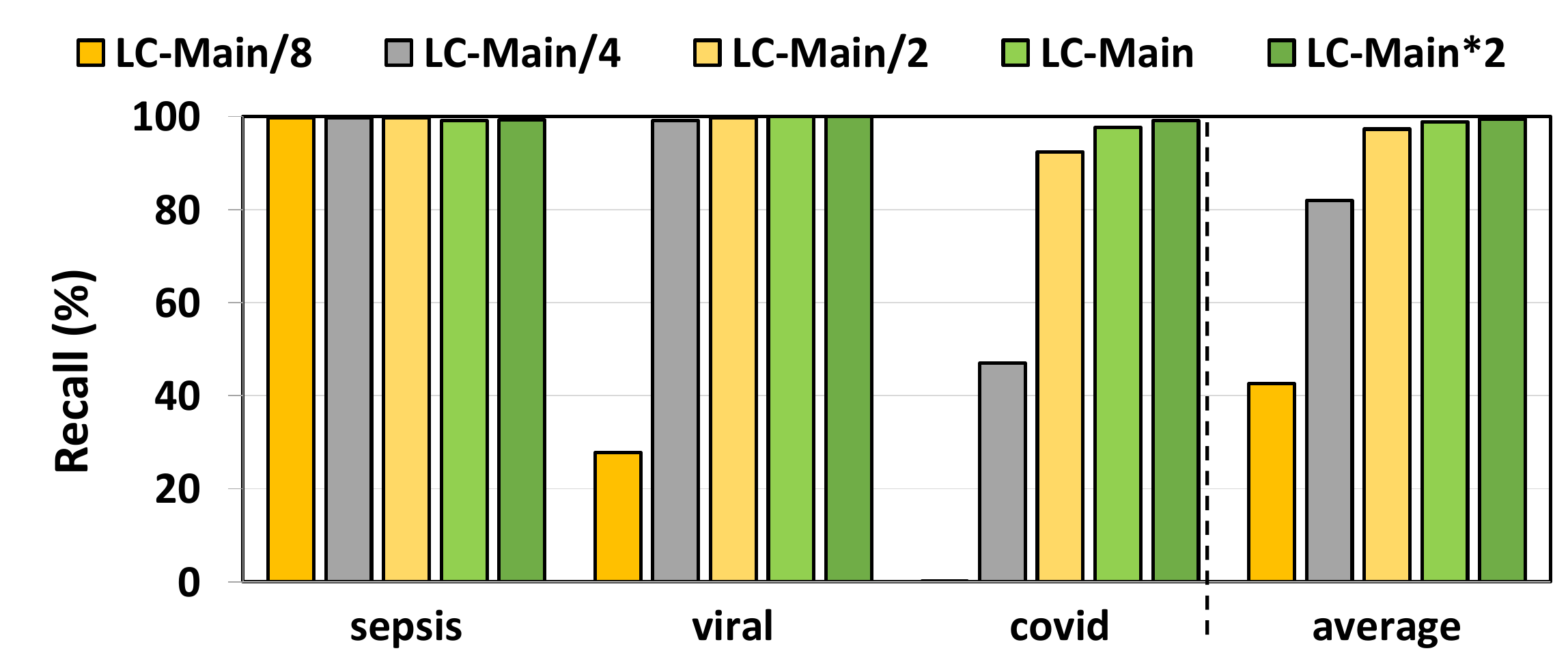}
  \caption{Recall {for our evaluated use-cases while using \proposal{} with different} \BaseConv{} configurations.}\vspace{-3pt}
  \label{fig:recall}
\end{figure}

In Figures~\ref{fig:precision} and ~\ref{fig:recall}, we assess the precision and recall of \proposal{} with different \BaseConv{} configurations for all use cases explained in Section~\ref{subsec:evalmethodology}, respectively. We make four key observations. First, \proposal{}'s precision and recall are between \worstprecision{}-\bestprecision{}~and \worstrecall{}-\bestrecall{} for different configurations of \BaseConv{} on average across all three use cases tested, respectively. Second, the precision and recall of \proposal{} increases as the model complexity of \BaseConv{} increases. Third, increasing the model complexity provides diminishing precision and recall improvements beyond the complexity of the $LC_{Main}$ model. Fourth, models smaller than $LC_{Main}$ are sufficient for use cases with small-to-medium target reference sizes, whereas more complex models are required for use cases with large target reference sizes. We conclude that $LC_{Main*2}$ provides the highest precision and highest recall compared to other \BaseConv{} configurations.

\subsection{Basecalling and Relative Abundance Accuracy} \label{sec:endaccuracy}
Table~\ref{tab:end-accuracy} shows the basecalling accuracy and the relative abundance (RA) deviation from the ground truth relative abundance estimation calculated for each \BaseConv{} configuration and using Bonito without pre-basecalling filtering.
We evaluate the RA accuracy of \proposal{} in the viral detection use case. We use Equation~(\ref{eq:1}) to calculate the RA deviation as the RA accuracy metric.

\begin{table}[h]
\centering
\caption{Basecalling Accuracy and Relative Abundance (RA) Deviation.}\label{tab:end-accuracy}
\begin{tabular}{@{}lcc@{}}\toprule
\textbf{Model} & \textbf{Basecalling} & \textbf{RA} \\
\textbf{Name} & \textbf{Accuracy} & \textbf{Deviation} \\\midrule
Bonito	&	94.60\% & 0.00\% \\\midrule
$LC_{Main\times2}$	 & 90.91\% & 0.03\% \\\midrule
$LC_{Main}$ &	89.75\%  & 0.08\% \\\midrule
$LC_{Main/2}$ & 86.83\%  & 0.23\% \\\midrule
$LC_{Main/4}$ &	80.82\%   & 0.91\% \\\midrule
$LC_{Main/8}$ &	70.42\%   & 72.19\% \\\midrule
\end{tabular}
\end{table}

We make the following two key observations. First, the RA deviation results are negligible ($\leq$0.1\%) for \proposal{} configurations with recall higher than 98.5\%. The only exception to this observation is our results when using the $LC_{Main/8}$ model. Although the basecalling accuracy drop is around 10\% between $LC_{Main/8}$ and $LC_{Main/4}$, the deviation increases substantially because most of the reads cannot be mapped (see the low recall result in Figure~\ref{fig:recall}), which affects the relative abundance estimations. Furthermore, as the read accuracy decreases, the read deviates from its original viral genome while it can still map to other viral genomes. We note that a similar issue is not observed in the precision results of the filtering use case (see Figure~\ref{fig:precision}) as the goal of the filtering is to differentiate a viral genome from a bacterial genome (Section~\ref{sec:readaccuracy}), rather than correctly mapping a read to its original viral genome compared to other viral genomes.

Second, \proposal{}'s minor inaccuracy is not biased toward any specific portion of the target reference. Otherwise, the deviation of the relative abundances would be higher. Losing a small number of \target{} reads randomly enables sequencing depth-of-coverage to compensate for the loss of reads. We conclude that \proposal{}'s high recall enables accurate estimation for relative abundance calculations.

\subsection{Basecalling Execution Time} \label{sec:performance}
\begin{figure}[h]
  \centering
  \includegraphics[width=0.95\linewidth]{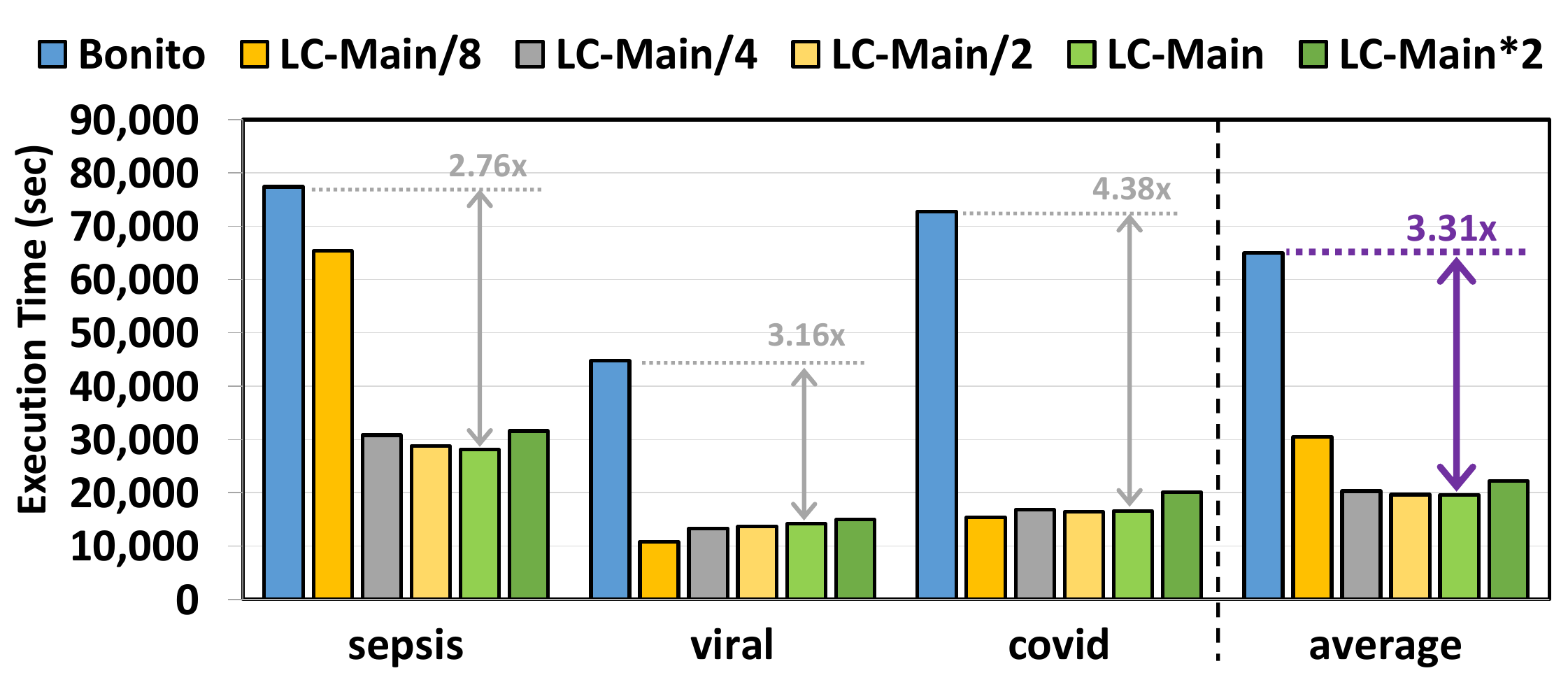}
  \caption{Basecalling execution time {for our evaluated use-cases while using \proposal{} with different} \BaseConv{} configurations.}\vspace{-3pt}
  \label{fig:perf}
\end{figure}

 Figure~\ref{fig:perf} provides the total execution time of Bonito and Bonito with \proposal{}. We make three key observations. First, \proposal{} improves runtime performance of Bonito by \worstperf{}-\bestperf{}. Second, both precision and model complexity of \BaseConv{} affect the runtime performance of \proposal{}, resulting in a non-linear relationship between model complexity and runtime performance. Precision affects the runtime performance of the filter as lower precision results in a higher number of falsely accepted reads to be basecalled using conventional basecallers. Model complexity affects the runtime performance of the filter as lower model complexity results in higher \BaseConv{} runtime performance with lower precision. Third, decreasing the model complexity increases the runtime performance up to the point where read filtering accuracy is no longer sufficient to filter out reads correctly. This results in a significant number of reads being falsely accepted by the filter, reducing the runtime performance of \proposal{} significantly. We conclude that \proposal{} significantly improves the execution time of basecalling.

\subsection{End-to-End Execution Time} 
\label{sec:endperformance}
\begin{figure}[h]
  \centering
  \includegraphics[width=0.95\linewidth]{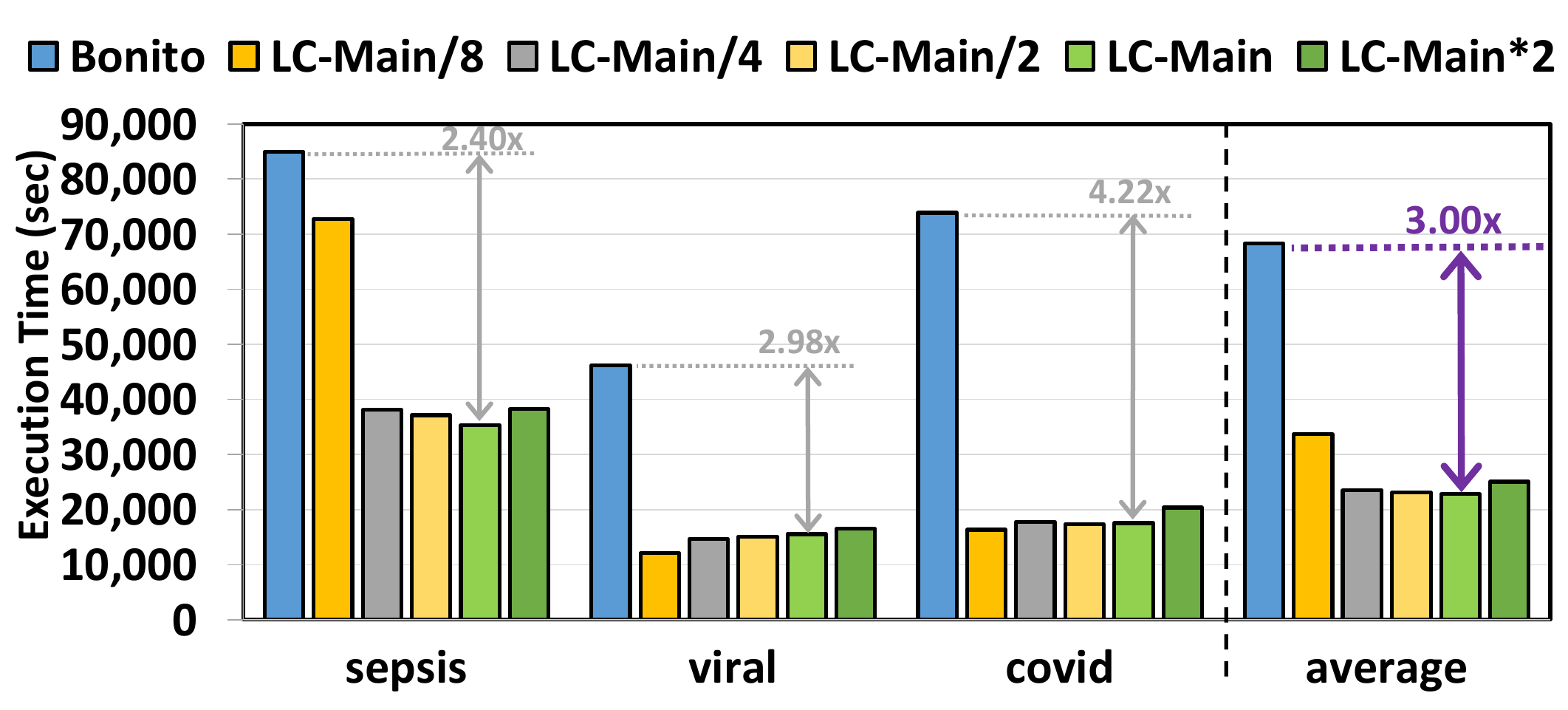}
  \caption{End-to-end execution time {for our evaluated use-cases while using \proposal{} with different} \BaseConv{} configurations.}\vspace{-3pt}
  \label{fig:endperf}
\end{figure}

 Figure~\ref{fig:endperf} provides the total time spent on the genome analysis pipeline of basecalling, read mapping, and variant calling with and without the use of pre-basecalling filtering. We used minimap2~\cite{li_minimap2_2018} and DeepVariant~\cite{poplin2018universal} for read mapping and variant calling, respectively. We make three key observations. First, \proposal{} improves the runtime performance of the entire genome sequence analysis pipeline by \worstendperf{}-\bestendperf{}. Second, the choice of the variant callers affects the end-to-end runtime performance improvement of \proposal{}. Since we used a highly accurate neural network-based variant caller, the execution time of variant calling dominated read mapping (not shown). This reduces the end-to-end runtime performance benefits of \proposal{}, as the variant calling is performed only on the alignments of \target{} reads (i.e., on the reduced dataset) determined during a relatively lightweight read mapping step. Third, similar to basecalling execution time, multiple factors affect the runtime performance of \proposal{} with different \BaseConv{} configurations, which results in a non-linear relationship between the runtime performance and model complexity of \BaseConv{}. We conclude that \proposal{} significantly improves the end-to-end execution time of the genome sequence analysis pipeline by providing up to \bestendperf{} speedup.

\subsection{Comparison to Prior Work}\label{subsec:comparepriorwork}
We compare \proposal{} with the best \BaseConv{} configuration (LC$_\text{Main}$) with two state-of-the-art adaptive sampling methods UNCALLED and Sigmap. Supplementary Section~\ref{subsec:bestmodel} explains the best model selection procedure. We used two different reference genomes for the sepsis use case, but UNCALLED failed to generate the index structure for our default human reference genome (hg38) in a high-end server with 1TB of main memory. We evaluate only these two methods, as they can readily be repurposed as pre-basecalling filters. The other methods are either not open-sourced fully~\cite{payne2020readfish} or cannot be repurposed as pre-basecalling filters as is~\cite{bao2021squigglenet}.

\begin{figure}[h]
  \centering
  \includegraphics[width=0.95\linewidth]{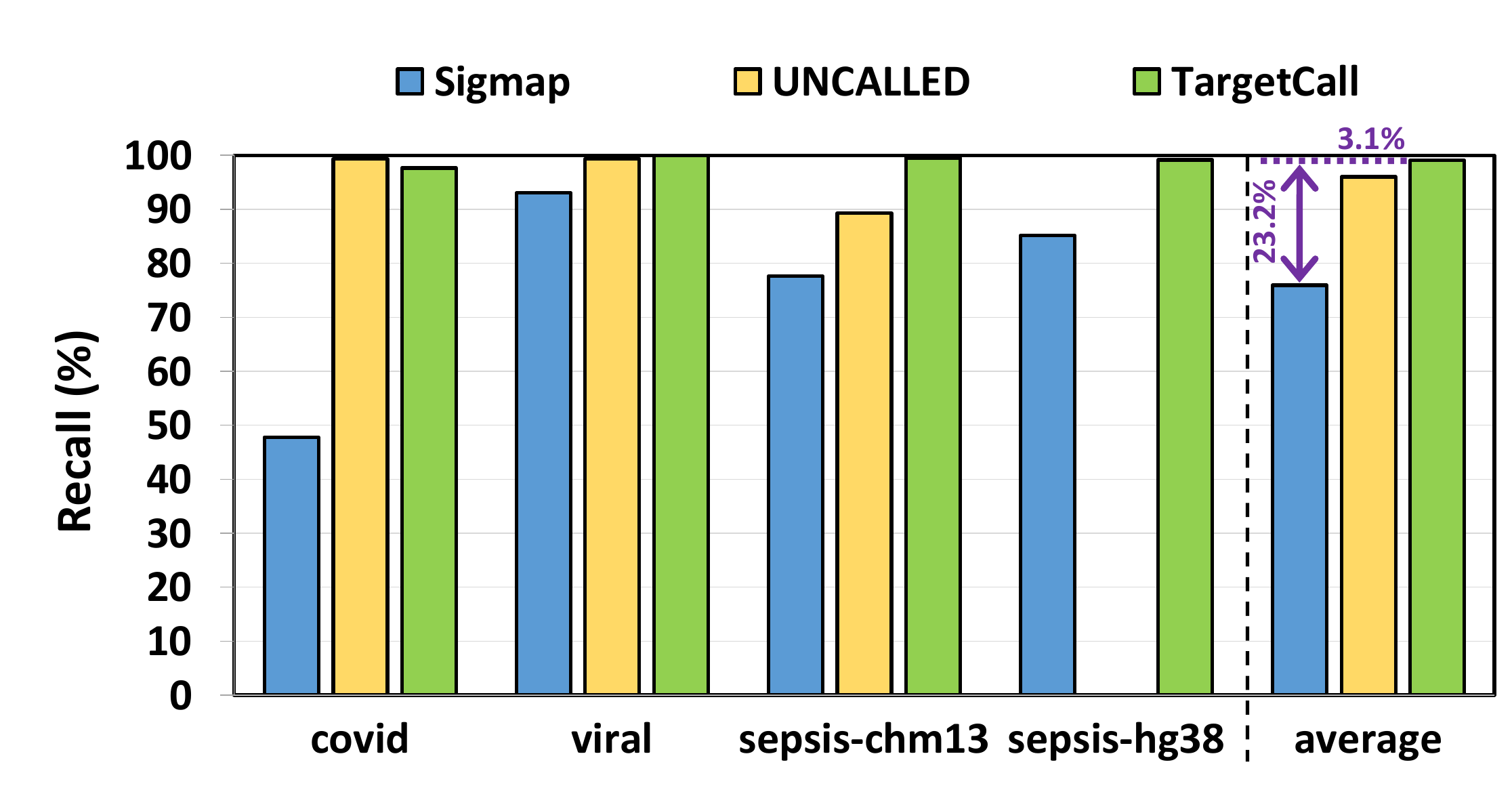}
  \caption{Recall of Sigmap, UNCALLED and \proposal{}}\vspace{-3pt}
  \label{fig:prior_recall}
\end{figure}

We compare the recall of \proposal{} with that of Sigmap and UNCALLED. The ground truth \target{} reads are determined by the conventional pipeline. In our analysis, we observe that UNCALLED could not be executed on the hg38 reference genome. To ensure a fair assessment, we excluded hg38 from the performance evaluation of UNCALLED. Figure~\ref{fig:prior_recall} shows the recall of Sigmap, UNCALLED and \proposal{}. We make two key observations. First, \proposal{} provides significantly higher recall, +\recallimpoveruncalled{}/+\recallimpoversigmap{}, than UNCALLED/Sigmap on average. Second, \proposal{} provides consistently the best recall than both methods across all use cases except covid detection, for which adaptive sampling methods are optimized. 

\begin{figure}[h]
  \centering
  \includegraphics[width=0.95\linewidth]{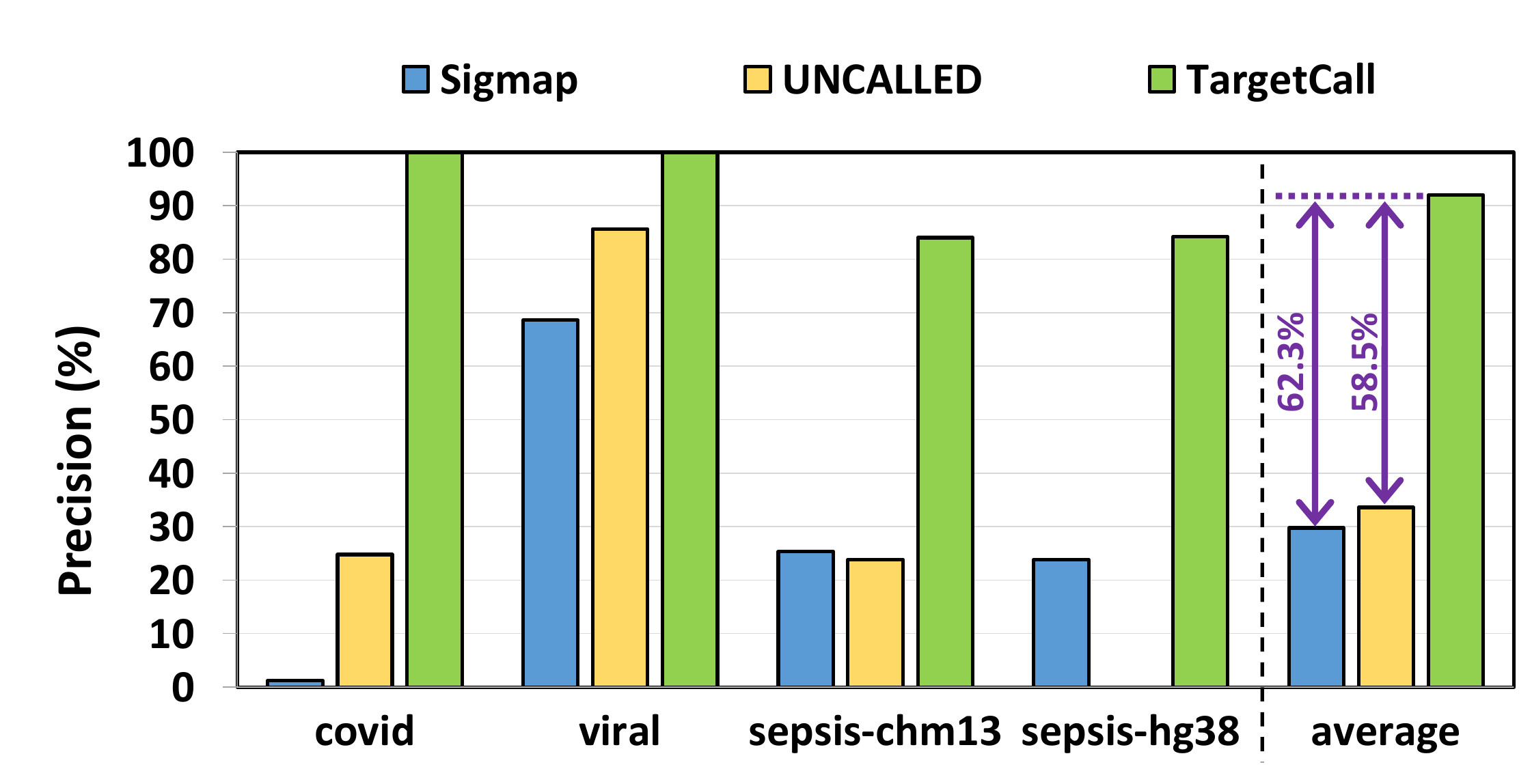}
  \caption{Precision of Sigmap, UNCALLED and \proposal{}}\vspace{-3pt}
  \label{fig:priorprecision}
\end{figure}

We compare the precision of \proposal{} with that of Sigmap and UNCALLED. Figure~\ref{fig:priorprecision} shows the precision of Sigmap, UNCALLED, and \proposal{}. We make two key observations. First, \proposal{} provides significantly higher precision, +\precisionimpoveruncalled{}/+\precisionimpoversigmap{}, than UNCALLED/Sigmap on average. Second, \proposal{} maintains high precision as the target reference size increases, unlike prior methods.  We conclude that \proposal{} significantly achieves higher recall and precision in all use cases independent of the target reference size.

\begin{figure}[h]
  \centering
  \includegraphics[width=\linewidth]{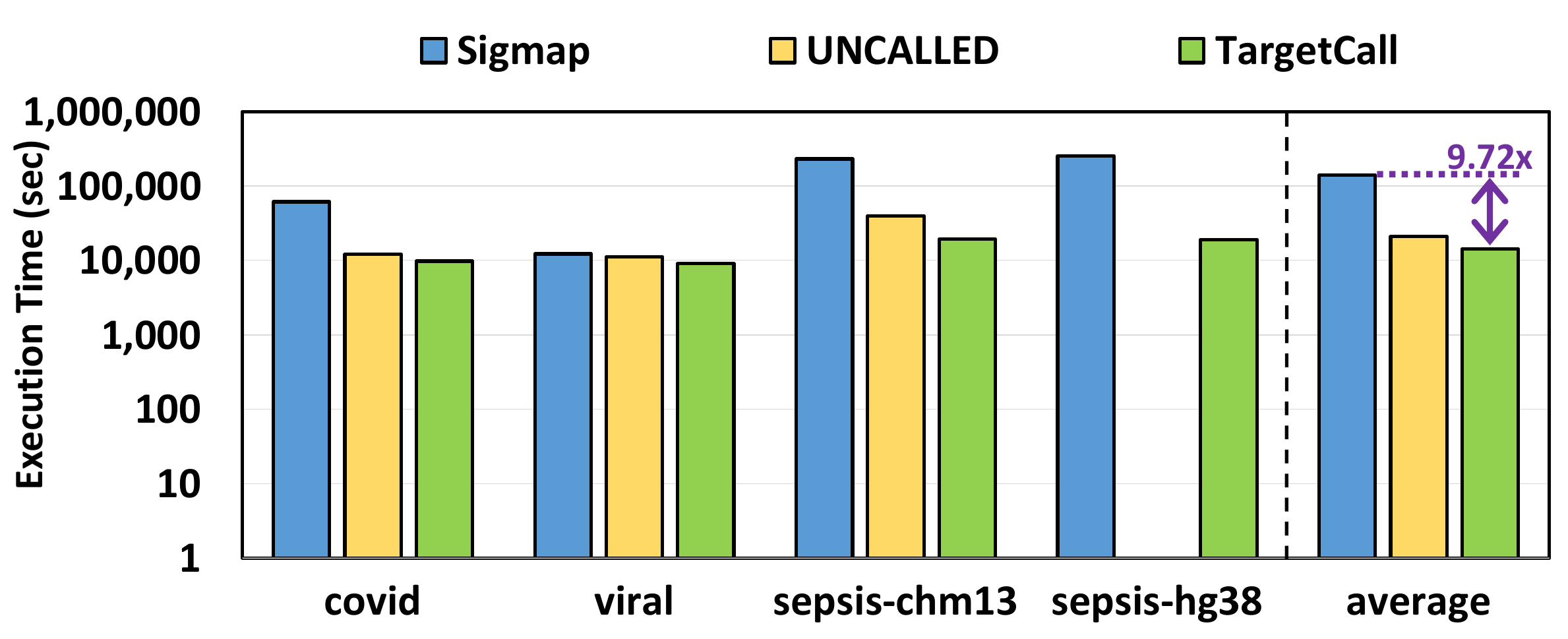}
  \caption{{End-to-end e}xecution {t}ime of Sigmap, UNCALLED{,} and \proposal{}}\vspace{-1pt}
  \label{fig:priorperf}
\end{figure}

We compare the end-to-end execution time of \proposal{} with that of Sigmap and UNCALLED, the results are shown in Figure~\ref{fig:priorperf}.  We make two key observations. First, we observe that \proposal{} outperforms Sigmap by \perfimpoversigmap{} and UNCALLED by \perfimpoveruncalled{}. \proposal{}'s higher runtime performance benefits come from its higher precision in filtering out \nontarget{} reads. Second, \proposal{}'s runtime performance improvements become more significant as the target reference size increases. 

\begin{figure}[h]
  \centering
  \includegraphics[width=\linewidth]{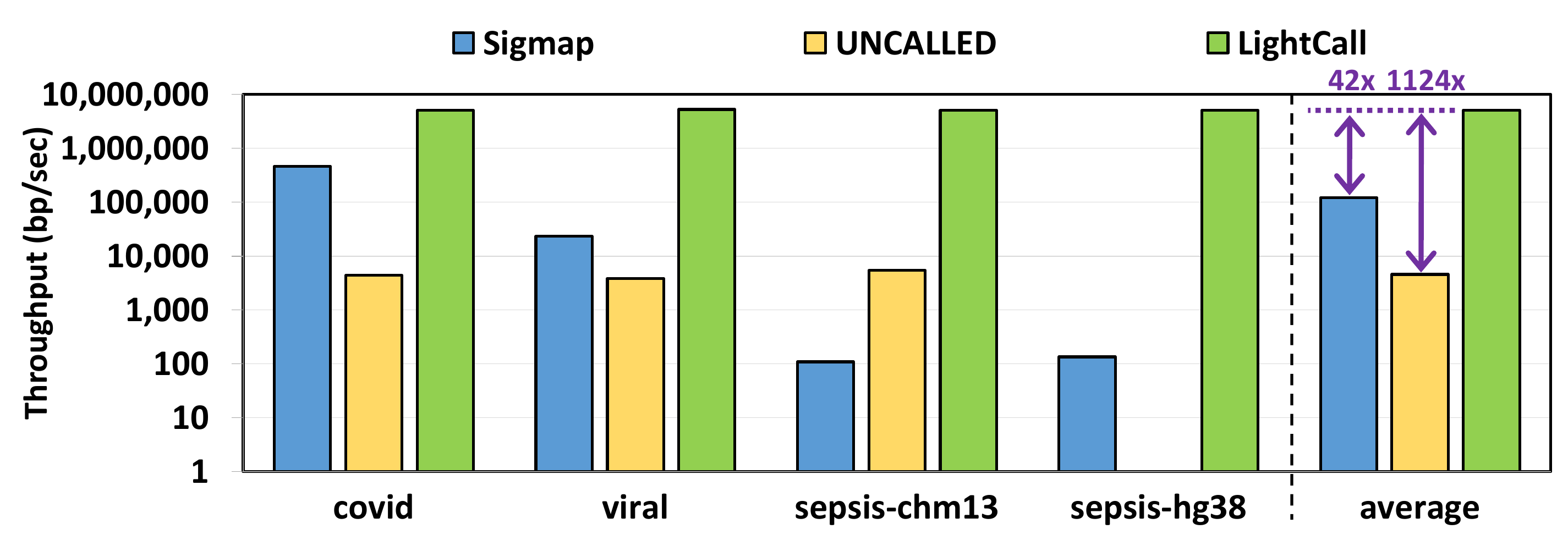}
  \caption{Throughput of Sigmap, UNCALLED{,} and \BaseConv{}}\vspace{-1pt}
  \label{fig:throughput}
\end{figure}

We evaluate the throughput of \proposal{} and compare it with that of Sigmap and UNCALLED. We evaluate throughput as the base pairs tools process per second. We use \texttt{uncalled pafstats} for evaluating the throughput of UNCALLED and Sigmap and bonito output for evaluating the throughput of \BaseConv{}. We only evaluate the throughput of \BaseConv{} component of \proposal{} as it has a significantly lower throughput than \SimCheck{} and is the bottleneck of \proposal{}. Figure~\ref{fig:throughput} shows the throughput of Sigmap, UNCALLED, and \BaseConv{}. We make three key observations. First, we observe that \BaseConv{} improves the throughput of UNCALLED/Sigmap by \throughputimpoveruncalled{}/\throughputimpoversigmap{}. Second, we observe that throughput of \BaseConv{} is consistently high unlike other tools (e.g., Sigmap) whose throughput reduce with target reference length. Third, \BaseConv{}'s high throughput does not reflect its execution time. The reason for this discrepancy between \BaseConv{}'s throughput and execution time is likely because, unlike Sigmap and UNCALLED, \proposal{} processes the entire read before labeling it as \target{}/\nontarget{}. We conclude that \proposal{}'s benefits can be amplified by optimizing it further via integrating early filtering of reads that will match/not match to the target reference without processing the entire read. 

Overall, we conclude that \proposal{} 1) significantly improves the end-to-end execution time of basecalling compared to prior methods by filtering out higher fraction of the \nontarget{} reads with its higher precision, 2) improves the recall of prior methods in filtering reads, 3) has a consistently higher throughput than all prior works, and 4) can be accurately applied to target reference lengths that prior methods are unable to be applied accurately while requiring much less (on average $5.76\times$, see Supplementary Section~\ref{subsec:peakmem}) peak memory compared to prior works. We believe \proposal{} can be used as a lightweight real-time sequence classification and filtering tool due to its high throughput if it is optimized to work with the first few chunks of a read. 

\subsection{\proposal{} Execution Time Breakdown}\label{subsec:breakdown}
\begin{figure}[h]
  \centering
  \includegraphics[width=0.7\linewidth]{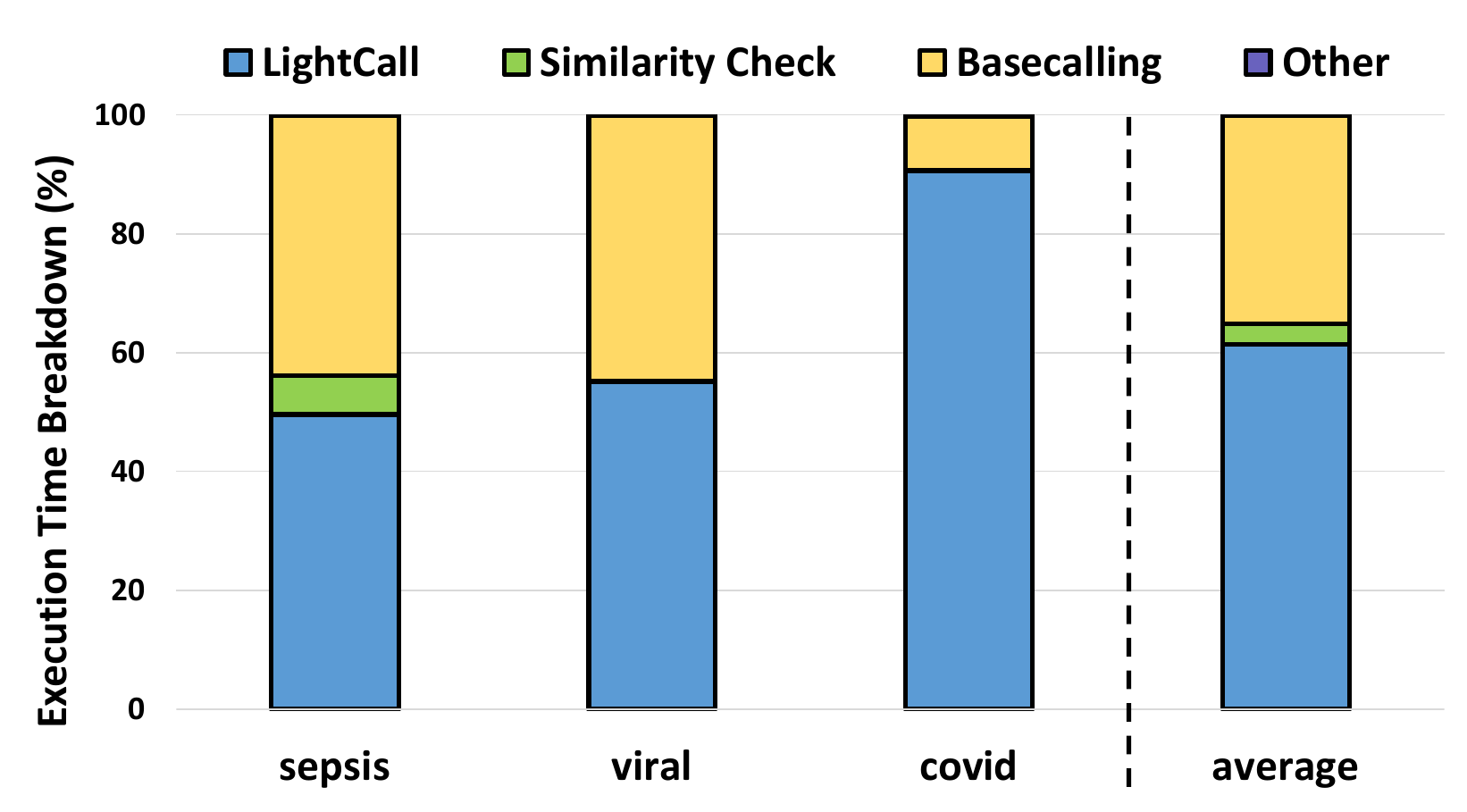}
  \caption{Execution Time Breakdown of \proposal{}}\vspace{-1pt}
  \label{fig:breakdown}
\end{figure}

We analyzed the execution time breakdown of a pipeline that includes \proposal{} as the pre-basecalling filter in Figure~\ref{fig:breakdown}. We make four key observations. First, \BaseConv{} is the bottleneck of the new basecalling pipeline that includes pre-basecalling filtering by consuming $61.44\%$ of the total execution time on average. Second, basecalling is still an important computational overhead for the pipeline by consuming $35.04\%$ of the total execution time on average. Third, the computational overhead of basecalling increases with the increased ratio of \target{} reads in the dataset, such as in the viral use case. Fourth, the \SimCheck{} component consumes less than $3.4\%$ of the total execution time on average and does not bottleneck the pipeline that includes \proposal{} as the pre-basecalling filter even when used in the expensive alignment mode. 

\section{Discussion} \label{sec:conclusion} 

We propose \proposal{}, a pre-basecalling filtering mechanism for eliminating the wasted computation in basecalling. \proposal{} performs light-weight basecalling to compute noisy reads using \BaseConv{}, and labels these noisy reads as \target{}/\nontarget{} using \SimCheck{}. \proposal{} eliminates the wasted computation in basecalling by performing basecalling only on the \target{} reads. We focus on convolution-based networks for \proposal{} architecture for two reasons: (a) matrix multiplication, the core operation in these networks, is highly suitable for hardware acceleration, facilitating improved performance, and (b) the training and inference of RNN and LSTM models involve sequential computation tasks, which pose significant challenges for acceleration on modern hardware such as GPUs and field-programmable gate arrays (FPGAs)~\cite{singh_rubicon_2024}. We evaluate \proposal{} for three different genome sequence analysis use cases of pre-basecalling filtering with varying requirements: covid detection, sepsis detection, and viral detection. We show that \proposal{} reduces the execution time of basecalling by filtering out the majority of the \nontarget{} reads and is more applicable compared to the state-of-the-art adaptive sampling methods. We hope that \proposal{} inspires future work in pre-basecalling filtering and real-time sequence classification that accelerate other bioinformatics workloads and emerging applications with its high throughput and recall. We explain future work and optimizations that can build upon \proposal{} in Supplementary Section~\ref{sec:optimizations}.

\section*{Acknowledgments}
SAFARI Research Group acknowledges the generous gift funding provided by our industrial partners (especially by Google, Huawei, Intel, Microsoft, VMware), which has been instrumental in enabling the 15+ year-long research SAFARI Research Group has been conducting on accelerating genome analysis. This work is also partially supported by the Semiconductor Research Corporation (SRC), the European Union’s Horizon programme for research and innovation [101047160 - BioPIM] and the Swiss National Science Foundation (SNSF) [200021\_213084].

\bibliographystyle{IEEEtran}
\setstretch{0.75}
{\small \bibliography{main}}

\onecolumn
\setcounter{secnumdepth}{3}
\clearpage
\begin{center}
\textbf{\LARGE Supplementary Material for\\ \ltitle}
\end{center}
\setcounter{section}{0}
\setcounter{equation}{0}
\setcounter{figure}{0}
\setcounter{table}{0}
\setcounter{page}{1}
\makeatletter
\renewcommand{\theequation}{S\arabic{equation}}
\renewcommand{\thetable}{S\arabic{table}}
\renewcommand{\thefigure}{S\arabic{figure}}
\renewcommand{\thesection}{\Alph{section}}
\renewcommand{\thesubsection}{\thesection.\arabic{subsection}}
\renewcommand{\thesubsubsection}{\thesubsection.\arabic{subsubsection}}

\newcommand{\TextUnderscore}{\rule{.4em}{.4pt}}

\section{Adaptive Sampling Background}
\label{sec:adaptivesampling}
Targeted sequencing is a method to selectively sequence reads from  the reference genome of interest (i.e., target reference) during sequencing time~\citesupp{supp_kovaka2021targeted, supp_gilpatrick2020targeted, supp_payne2020readfish}. {ONT devices have the potential to enable computational targeted sequencing without the need for library preparation with a feature, known as} \emph{Read Until}~\citesupp{supp_kovaka2021targeted, supp_loose2016realtime}. ONT sequencers that support Read Until can selectively remove a read from the nanopore while the read is {being} sequenced, called as the adaptive sampling. Read Until can be used to remove \nontarget{} reads during sequencing, eliminating wasted computation on \nontarget{} reads in basecalling. All adaptive sampling approaches work by labeling a read as \target{} or \nontarget{} and stopping the sequencing of \nontarget{} reads immediately after labeling using Read Until. {We discuss three groups of works that perform Read Until}, classified based on their methodology to label the read. The first group converts the target reference into a reference raw signal and performs raw signal-level alignment~\citesupp{supp_zhang2021realtime, supp_dunn2021squigglefilter, supp_loose2016realtime}. The second group generates noisy sequence representations of the raw signal to compare them with the target reference~\citesupp{supp_kovaka2021targeted, supp_payne2020readfish}. The {third} group of works utilize neural network classifiers to label the sequences~\citesupp{supp_bao2021squigglenet, supp_noordijk_baseless_2023}. To our knowledge, none of these works can be fully repurposed as widely-applicable pre-basecalling filters to eliminate the wasted computation in basecalling for a wide range of genome applications as their accuracy drops significantly with the increasing target reference size.
  
{The first group of work compares the raw electrical signals to a target reference without basecalling the signals in two steps~\citesupp{supp_loose2016realtime}. The first step converts the bases of the reference genome into their \emph{synthetic} raw signal representation. The second step identifies similarities between the raw read signal and the synthetic reference signal. Prior work~\citesupp{supp_loose2016realtime} used the Dynamic Time Warping (DTW) algorithm that measures the similarity of a pair of signals to find the similarity. If the read is not similar to} target reference, it is labeled as \nontarget{}. Unlike \proposal{}, this method {is not applicable to large target references of millions of base pairs (bp) long due the quadratic time complexity of DTW with respect to the length of the sequences.} To address the computational bottleneck of DTW, Sigmap~\citesupp{supp_zhang2021realtime} propose to generate an index for the synthetic reference signal. Sigmap queries the generated index as the reads are being sequenced to find the potential similarity positions between the read and the reference signal, avoiding the DTW calculation. These positions refer to matches of short subsequences of the read and the reference signal, and the final labelling is determined based on chaining these match locations. The use of index structure enables Sigmap to be applicable to target references of length up to $\sim$100 Mbp, however, Sigmap is still significantly less applicable compared to \proposal{}. Recently, SquiggleFilter proposed to design an accelerator to make DTW calculation faster~\citesupp{supp_dunn2021squigglefilter} but it was unable to make DTW scalable for large target references.

{The second group of works is} based on converting the raw signal into a set of bases and comparing these bases to the reference to label the raw signal~\citesupp{supp_payne2020readfish, supp_kovaka2021targeted}. Readfish uses a real-time basecalling method to basecall the read as it is being sequenced and perform read mapping on the basecalled portion of the raw signal~\citesupp{supp_payne2020readfish}. Since basecallers are optimized to work on complete reads, this method results in suboptimal base sequences hence may incorrectly label the reads~\citesupp{supp_zhang2021realtime}. To mitigate these drawbacks, UNCALLED uses an index of the reference genome to probabilistically convert the raw signal into a set of short nucleotide subsequences called seeds and cluster them~\citesupp{supp_kovaka2021targeted}. The read is classified as \target{} read if there is a location in the target reference that has significantly more seeds mapping to that region than the others~\citesupp{supp_kovaka2021targeted}. UNCALLED is also designed to work \emph{only} on small reference genomes with the goal of performing adaptive sampling~\citesupp{supp_zhang2021realtime, supp_kovaka2021targeted}.

{The third group of works uses machine learning to label the raw signals in a sample without performing basecalling and costly analyses in the base space~\citesupp{supp_bao2021squigglenet, supp_noordijk_baseless_2023}. SquiggleNet~\citesupp{supp_bao2021squigglenet} can identify a certain class of species with a high accuracy where the class membership is determined based on the target reference. However, this approach requires training the machine learning model for each target reference, which cannot be practically applied to classify any type of species due to the high computational costs of training. Therefore, unlike \proposal{}, SquiggleNet cannot be used as a widely-applicable pre-basecalling filter. BaseLess~\citesupp{supp_noordijk_baseless_2023} utilizes an array of small neural networks to detect a small subsequences from raw signals and match these subsequences with a target reference that share the same subsequence. This design choice provides a flexible solution that can define the set of pre-trained neural network models of subsequences to identify a certain target reference instead of retraining the neural network model for each species. Unfortunately, none of these works can avoid the cost of training the models multiple times (i.e., for each subsequence or species) to identify the target references in raw signals.}

\heads{Limitations of Adaptive Sampling Approaches}
Even though these approaches can discard \nontarget{} reads from the genome analysis pipeline hence eliminating the wasted computation in basecalling, all have at least one of the following three key limitations, preventing them to be used as \emph{widely applicable} pre-basecalling filters.
First, some~\citesupp{supp_kovaka2021targeted, supp_zhang2021realtime, supp_payne2020readfish} have low (77.5\%-90.40\%) recall which affects the accuracy of downstream analysis. These tools falsely reject a significant portion ($\sim$10\%-$\sim$23\%) of the \target{} reads. Second, some~\citesupp{supp_kovaka2021targeted, supp_zhang2021realtime, supp_dunn2021squigglefilter, supp_loose2016realtime} are not scalable to the long target references. This happens due to one of the following reasons: 1) use of signal-signal alignment algorithms whose complexity increases linearly with the target reference length~\citesupp{supp_dunn2021squigglefilter, supp_loose2016realtime}, 2) use of complex data structures to represent target reference in signal domain that are not scalable to long target reference lengths~\citesupp{supp_zhang2021realtime}, and 3) use of probabilistic algorithms that scale poorly with the increasing target reference length~\citesupp{supp_kovaka2021targeted}. These tools cannot be used for applications that require long target references, such as human reference. Third, some require neural network classifiers to be re-trained for each different application and target reference~\citesupp{supp_bao2021squigglenet}. The re-training is required since the classifier is trained depending on the specific set of \target{} and \nontarget{} reads. These approaches cannot be used in a wide range of pre-basecalling filtering applications without significant overheads.

\section{Evaluated Datasets}
\label{sec:dataset}
\heads{Evaluated Read Datasets}
We use four real and one simulated dataset to evaluate \proposal{}. Table~\ref{tab:dataset} {provides} details {on the evaluated} datasets. Datasets D1 \& D2, D1 \& D4 and D3 \& D5 are used for covid detection, sepsis detection and viral detection use cases respectively. For real datasets we randomly sample the datasets provided by prior research to keep a tractable experiment time. We use DeepSimulator to {generate} the simulated reads~\citesupp{supp_li2018deepsimulator, supp_li2020deepsimulator}. We simulate the dataset D5 due to unavailability of open access raw signal files for viral reads. 

\begin{table}[h]
\centering
\caption{Evaluated read datasets.}\label{tab:dataset}
\vspace{1mm}
\resizebox{0.7\columnwidth}{!}{
\begin{tabular}{@{}lcccc@{}}\toprule
\textbf{List of} & \textbf{ Dataset } & \textbf{Number of} & \textbf{Source} & \textbf{DOI} \\
\textbf{Read Datasets} & \textbf{Type} & \textbf{Reads} & & \textbf{Accession} \\\midrule
(D1) Human & Real	& 196,000 &  \citesupp{supp_zook2019open} &  10.5281/zenodo.7334648 \\ \cmidrule{1-5}
(D2) SARS-CoV-2	& Real	& 4,000 & \citesupp{supp_cadde} & 10.5281/zenodo.7335539 \\ \cmidrule{1-5}
(D3) Bacterial Mixture (n=7)	& Real	& 72,567 & \citesupp{supp_wick2019performance} & 10.5281/zenodo.7335525 \\ \cmidrule{1-5}
(D4) Bacterial Mixture (n=9)	& Real	& 15,200 &  \citesupp{supp_wick2019performance} &  10.5281/zenodo.7335517 \\ \cmidrule{1-5}
(D5) Viral Mixture (n=7)	& Simulated	& 35,000 & \citesupp{supp_li2018deepsimulator, supp_li2020deepsimulator} & 10.5281/zenodo.7334592 \\ \cmidrule{1-5}
\multicolumn{3}{l}{}\\ 
\end{tabular}}
\end{table}

\heads{Evaluated Reference Genomes}
 We use four reference genomes to evaluate \proposal{} on three applications. Table~\ref{tab:refgenomes} lists the details of {these} reference genomes. 

\begin{table}[h]
\centering
\caption{List of Reference Genomes Used for each Use Case.}\label{tab:refgenomes}
\vspace{1mm}
\resizebox{0.7\columnwidth}{!}{
\begin{tabular}{@{}lcccc@{}}\toprule
\textbf{Reference} & \textbf{Reference} & \textbf{Use} & \textbf{Datasets} & \textbf{\% of On-Target} \\
\textbf{Genome} & \textbf{Length (bp)} & \textbf{Case} & \textbf{Compared} & \textbf{ Reads} 
\\\midrule
Human (GRCh38) & 3,088,286,401 & Sepsis Detection & D1 \& D4 & 17.64\% \\  \cmidrule{1-5}
Human (Chm13) & 3,117,292,070 & Sepsis Detection & D1 \& D4 & 17.64\% \\  \cmidrule{1-5}
SARS-CoV-2 & 29,903	& Covid Detection & D1 \& D2& 2\% \\ 
(NC\_045512.2) & 	&   &  & \\ 
\cmidrule{1-5}
Viral Combined (n=10) & 861,552 & Viral Detection & D3 \& D5 & 32.51\% \\  \cmidrule{1-5}
\end{tabular}}
\end{table}

For testing viral detection use case, we used 10 viral reference genomes from NCBI RefSeq (with IDs NC\_003977.2, AC\_000007.1, NC\_009334.1, NC\_001526.4, NC\_010277.2, NC\_001731.1, NC\_063383.1, NC\_055231.1, NC\_045512.2, NC\_014361.1); and combined them to use as the target reference~\citesupp{supp_refseq}. Only 7 of these 10 reference genomes are used for simulating reads of dataset D5\footnote{Reference Genomes tested can be accessed with DOI 10.5281/zenodo.7335545}.

\section{Results} \label{sec:results}

\subsection{Best Model Selection}\label{subsec:bestmodel}

\begin{figure}[h]
  \centering
  \includegraphics[width=0.6\linewidth]{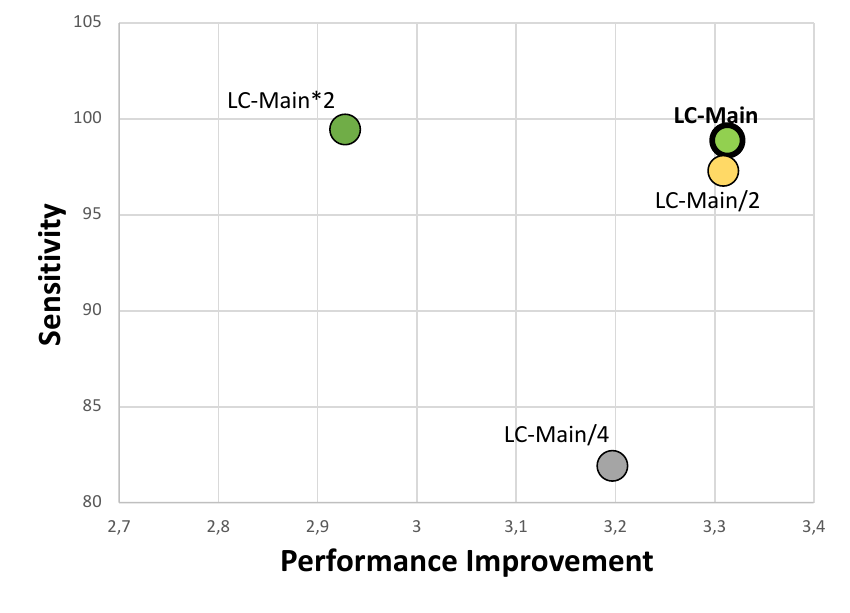}
  \vspace{-6pt}
  \caption{{Overall recall and performance (basecalling speedup) improvement for \proposal{} with different \BaseConv{} configurations}}\vspace{-3pt}
  \label{fig:bestmodel}
\end{figure}

We evaluate the performance-recall trade-off of \proposal{} to determine the best \BaseConv{} architecture for \proposal{}. Figure~\ref{fig:bestmodel} aggregates the recall and performance improvement (i.e., basecalling speedup) of all \BaseConv{} configurations evaluated (except $LC_{Main/8}$). We make the following three key observations. First, $LC_{Main}$ provides the highest (\perfoverbasecalling{}) speedup in basecalling. Second, $LC_{Main*2}$ provides the highest (\bestrecall{}) recall in basecalling. Third, $LC_{Main}$ provides significantly higher (13.36\%) speedup than $LC_{Main*2}$ with minimal (0.57\%) reductions in recall. Therefore, we select $LC_{Main}$ as the \BaseConv{} component of \proposal{}. {We conclude that} \proposal{} { by using $LC_{Main}$} improves the performance of basecalling by \perfoverbasecalling{} by precisely filtering \precisionoverbasecalling{} of the \target{} reads while maintaining high (\accoverbasecalling{}) recall in filtering.

\subsection{Peak Memory}\label{subsec:peakmem}

We evaluate the peak memory of \proposal{} and compare it against with that of Sigmap and UNCALLED. We subsample $4000$ human reads from the dataset to evaluate the peak memory usage of these tools since adding more reads to the experiment does not change the results. We use the linux time command for evaluating the peak memory usage of UNCALLED, Sigmap and \SimCheck{}. We use the nvidia-smi command to evaluate the peak memory usage of \BaseConv{}. The peak memory usage of \proposal{} is equal to the maximum peak memory usage of its components: \BaseConv{} and \SimCheck{}. Table~\ref{tab:peakmemory} demonstrates the peak memory usage of Sigmap, UNCALLED, \proposal{} and its components. We make four observations. First, \BaseConv{}'s peak memory usage is the same across different use cases, since the basecalling algorithm is independent of the target reference. Second, for use cases with small target reference, the peak memory usage of \proposal{} is determined by the peak memory usage of \BaseConv{}, whereas its peak memory usage is determined by the peak memory usage of \SimCheck{} for large target references. Third, \proposal{}'s peak memory usage is always less than Sigmap's peak memory usage, and it becomes more efficient in terms of peak memory compared to Sigmap as the target reference length increases. Fourth, both UNCALLED's and \proposal{}'s peak memory usage are low for all use cases ($\leq 20$GB). We conclude that \proposal{} is the only tool that both has low peak memory usage for all target reference lengths and is applicable to all target references. 

\begin{table}[h]
\centering
\caption{Peak Memory (MB).}\label{tab:peakmemory}
\vspace{1mm}
\resizebox{0.7\columnwidth}{!}{
\begin{tabular}{@{}lrrrrr@{}}\toprule
\textbf{Use Case} & \textbf{UNCALLED} & \textbf{Sigmap} & \textbf{\BaseConv{}} & \textbf{\SimCheck{}} & \textbf{\proposal{}} \\ \cmidrule{1-6}
Covid & 1329.1 & 3455.0 & 2005.0 &	139.6	& 2005.0 \\  \cmidrule{1-6}
Viral & 1385.4 & 	3546.8 &	2005.0 &	274.9 &	2005.0 \\  \cmidrule{1-6}
Sepsis\_chm13 & 8126.4 &	234504.3 &	2005.0 &	16593.9 &	16593.9 \\ \cmidrule{1-6}
Sepsis\_hg38 & - &	225746.5	& 2005.0 &	16974.7 &	16974.7 \\  \cmidrule{1-6}
\end{tabular}}
\end{table}

\section{Future Work and Optimizations} \label{sec:optimizations}
The performance and accuracy of \proposal{} can be further optimized in three main directions. The first direction is to optimize the \BaseConv{} component to achieve higher basecalling accuracy with significant performance improvements. Currently, the \BaseConv{} model is designed based on pruning a state-of-the-art basecaller, Bonito's model. The specific model configurations evaluated are chosen based on the insight we developed from prior work. We believe the accuracy and performance of \BaseConv{} can be optimized even better with a more methodological approach in specifying the precise configurations, such as neural architecture search~\citesupp{supp_zoph2017neural}. 

The second direction is to optimize the \SimCheck{} component to achieve higher recall, precision and/or performance. Currently, the state-of-the-art read mapper, minimap2, is used as the \SimCheck{} component of \proposal{}. \proposal{}'s recall can be improved by manually tuning the minimap2 parameters, and its performance can be improved by using it with large window size or without the expensive alignment mode. We left this analysis as part of future work for the following three reasons. First, although the recall and precision of \proposal{} can be improved by changing the minimap2 parameters, the \proposal{}’s precision and recall are already very high (92.1\%/99.1\% averaged over 4 use cases in Section 3.6). Second, as we show in section 3.7 of the paper that minimap2 consumes less than 3.4\% of the execution time of a basecalling pipeline that includes \proposal{} as the pre-basecalling filter. Therefore, the performance of \proposal{} will not be significantly improved by optimizing the minimap2 parameters. Third, changing the minimap2 parameters will change the reads labeled as \target{}, so the reads that needed to be basecalled using state-of-the-art basecallers to evaluate the execution time of \proposal{} in addition to its precision/recall. This will increase the experimental time extensively. Therefore, we left this optimization as part of our future work considering the already high recall, precision and performance of \proposal{}.

The third direction is to replace the \SimCheck{} component with a less expensive filter. Read mapping problem aims to match a read to its position in a reference genome. However, the goal of \SimCheck{}, predicting if a read is coming from a target reference, is simpler than read mapping. Therefore, \proposal{} can be optimized by using k-mer or sketch based methods as its \SimCheck{} component such as KrakenUnique~\citesupp{supp_krakenuniq2018breitwieser} and Mash~\citesupp{supp_mash2016ondov}. Similar to the second direction, we left this as part of future work, as \proposal{} is not bottlenecked by its \SimCheck{} component.

We hope that \proposal{} inspires future work in pre-basecalling filtering and raw signal classification that accelerate other bioinformatics workloads and emerging applications with its high throughput and recall. One such application is adaptive sampling where the raw signals are classified in real-time while the reads are being sequenced. We show in Section 3.6 of the paper that \proposal{}'s throughput is on par with sequencing throughput of ONT devices, and much higher (up to 1124x) than the adaptive sampling approaches we compared against. Further analysis is required to understand to what extent \proposal{} can be used for adaptive sampling such as evaluating the recall and precision of \proposal{} when only the initial portions of the raw signals are available for classification. We believe the optimizations explained above can be useful for future work that applies \proposal{} to adaptive sampling problem.

\let\noopsort\undefined
\let\printfirst\undefined
\let\singleletter\undefined
\let\switchargs\undefined

\bibliographystylesupp{IEEEtran}
\setstretch{0.75}
{\small \bibliographysupp{supp}}

\end{document}